\documentclass[sigplan,10pt]{acmart}
\AtBeginDocument{%
  }

\usepackage{algorithm}
\usepackage{algpseudocode}

\usepackage{booktabs} 
\usepackage{array}    
\usepackage{graphicx}
\usepackage{subcaption}
\usepackage{xspace}

\usepackage{booktabs} 
\usepackage{array} 
\usepackage{ragged2e} 

\algtext*{EndWhile} 
\algtext*{EndIf}    
\algtext*{EndFor}   

\usepackage{multirow}
\setcopyright{acmlicensed}
\copyrightyear{2018}
\acmYear{2018}
\acmDOI{XXXXXXX.XXXXXXX}
\acmConference[Conference acronym 'XX]{Make sure to enter the correct
  conference title from your rights confirmation email}{June 03--05,
  2018}{Woodstock, NY}
\acmISBN{978-1-4503-XXXX-X/2018/06}

\newcommand{\stitle}[1]{\noindent{\bf #1.\/}}

\newcommand{\squishlist}{
	\begin{list}{$\bullet$}
		{ \setlength{\itemsep}{1pt}
			\setlength{\parsep}{1pt}
			\setlength{\topsep}{2.5pt}
			\setlength{\partopsep}{0.5pt}
			\setlength{\leftmargin}{1em}
			\setlength{\labelwidth}{1em}
			\setlength{\labelsep}{0.6em}
		}
	}
	\newcommand{\squishend}{
	\end{list}
}

\begin{document}

\newcommand{\sys}{CoTra\xspace}
\title{Towards Efficient and Scalable Distributed Vector Search with RDMA}


\author{Xiangyu Zhi}
\affiliation{%
  \institution{The Chinese University of Hong Kong}
  \city{}
  \country{}}
\email{xyzhi24@cse.cuhk.edu.hk}

\author{Meng Chen}
\authornote{Work partially performed during the internship at Microsoft.}
\affiliation{%
  \institution{Fudan University}
  \city{}
  \country{}}
\email{mengchen22@m.fudan.edu.cn}

\author{Xiao Yan}
\affiliation{%
  \institution{Wuhan University}
  \city{}
  \country{}}
\email{yanxiaosunny@whu.edu.cn}

\author{Baotong Lu}
\affiliation{%
  \institution{Microsoft Research}
  \city{}
  \country{}}
\email{baotonglu@microsoft.com}

\author{Hui Li}
\affiliation{%
  \institution{The Chinese University of Hong Kong}
  \city{}
  \country{}}
\email{hli@cse.cuhk.edu.hk}

\author{Qianxi Zhang}
\affiliation{%
  \institution{Microsoft Research}
  \city{}
  \country{}}
\email{qianxi.zhang@microsoft.com}

\author{Qi Chen}
\affiliation{%
  \institution{Microsoft Research}
  \city{}
  \country{}}
\email{cheqi@microsoft.com}

\author{James Cheng}
\affiliation{%
  \institution{The Chinese University of Hong Kong}
  \city{}
  \country{}}
\email{jcheng@cse.cuhk.edu.hk}

\renewcommand{\shortauthors}{Trovato et al.}


\begin{abstract}

Similarity-based vector search facilitates many important applications such as search and recommendation but is limited by the memory capacity and bandwidth of a single machine due to large datasets and intensive data read. In this paper, we present \sys, a system that scales up vector search for distributed execution. We observe a tension between \textit{computation and communication} efficiency, which is the main challenge for good scalability, i.e., handling the local vectors on each machine \textit{independently} blows up computation as the pruning power of vector index is not fully utilized, while running a \textit{global index} over all machines introduces rich \textit{data dependencies} and thus extensive communication. To resolve such tension, we leverage the fact that vector search is approximate in nature and robust to \textit{asynchronous execution}. In particular, we run collaborative vector search over the machines with algorithm-system co-designs including clustering-based data partitioning to reduce communication, asynchronous execution to avoid communication stall, and task push to reduce network traffic. To make collaborative search efficient, we introduce a suite of system optimizations including task scheduling, communication batching, and storage format. We evaluate \sys on real datasets and compare with four baselines. The results show that when using 16 machines, the query throughput of \sys scales to 9.8-13.4$\times$ over a single machine and is 2.12-3.58$\times$ of the best-performing baseline at recall@10$\ge$0.95.

\end{abstract}





\maketitle

\section{Introduction}



\begin{figure}[!t]
  \centering
  \includegraphics[width=0.95\linewidth]{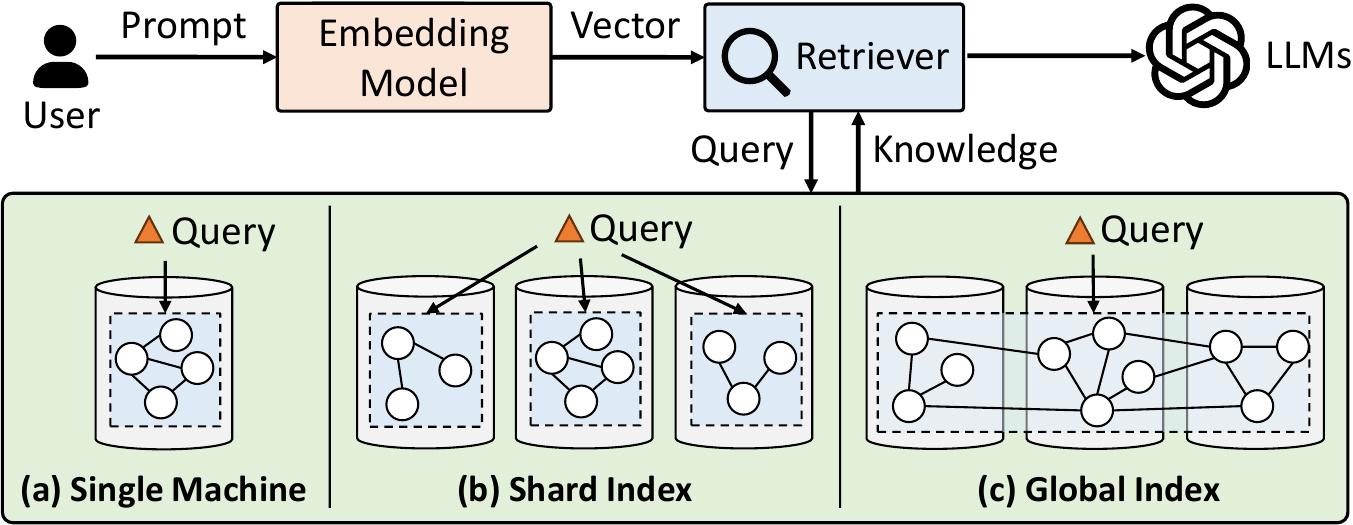}
  \caption{Vector search in RAG (top) and three designs of large-scale vector search (bottom).}
  \label{fig:intro:app}
\end{figure}

High-dimensional vector search is fundamental to many modern applications as it enables semantic similarity comparisons across diverse data types from text and images to genomic sequences. Specifically, complex data objects are mapped to vector embeddings~\cite{chen2024bge,radford2021learning,chen2024genept,zhao2021learning}, and then similarity-based \textit{vector search} is conducted to retrieve top-$k$ most similar vectors (e.g., measured by Euclidean distance) to a query vector~\cite{li2019approximate}. Since exact vector search requires a linear scan of high-dimensional vectors in a vector dataset, \textit{approximate vector search} has been widely adopted to retrieve most (instead of all) of the top-$k$ similar vectors to trade for efficiency. Approximate vector search powers numerous critical tasks like content-based search (e.g., for images and videos)~\cite{kalantidis2014locally,jegou2010product,zhang2018visual}, recommender systems (e.g., for e-commerce products and online advertisements)~\cite{van2016learning,nigam2019semantic,spann,huang2020embedding}, and bioinformatics~\cite{schutze2022nearest,bittremieux2018fast,kiselev2018scmap}. Figure~\ref{fig:intro:app} shows a popular application, retrieval-argumented generation (RAG) for LLMs~\cite{zhao2024retrieval}, where text chunks are embedded as vectors and user prompts act as queries, and retrieved text chunks provide external knowledge to LLMs for response generation.

The state-of-the-art index for vector search is proximity graph~\cite{li2019approximate,wang2021comprehensive}. Although there are many variants~\cite{hnsw,nsg,peng2023efficient}, the core idea is to organize vectors into a graph, where each vector is a node and connects to its similar vectors\footnote{We use ``node'' and ``vector'' interchangeably.}. Vector search is processed by a \textit{graph traversal}, which starts from a fixed or random entry node, computes similarities for all neighbors when visiting a node, and maintains a \textit{candidate queue} to record the computed similarities and choose the most similar but unvisited node as the next node to visit. 
For a dataset with $N$ vectors, the number of similarity computations required by the graph traversal is $\log N$ in scale~\cite{hnsw,nsg,prokhorenkova2020graph}.

\begin{figure}[!t]
  \centering
  \includegraphics[width=0.85\linewidth]{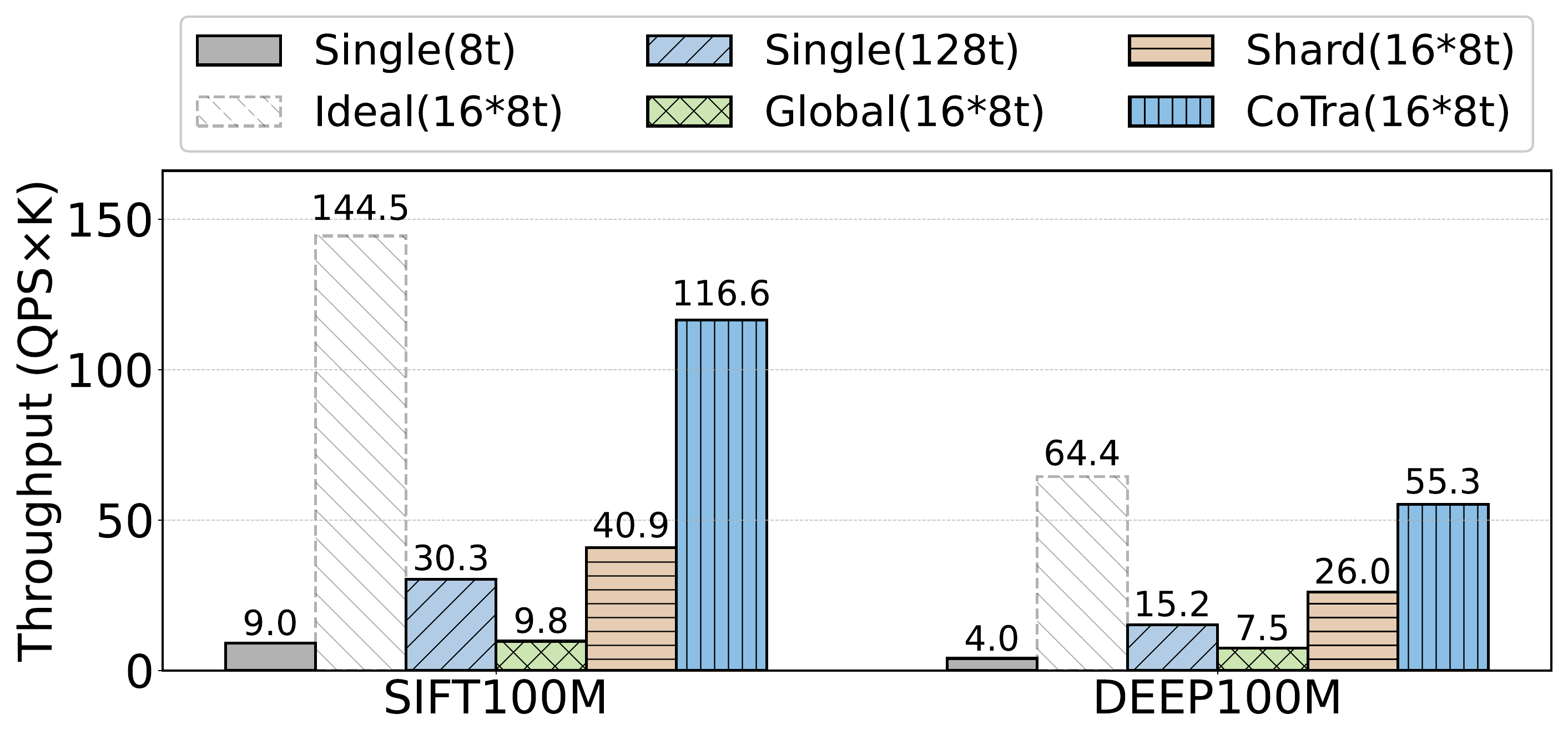}
  \caption{Query throughput (QPS) at 0.95 recall for top-10 vector search. Single-machine experiments use 8 and 128 threads; distributed solutions (\textit{Shard}, \textit{Global}, and our \textit{CoTra}) use 16 machines (8 threads each). \textit{Ideal} means ideal linear scaling for 16 machines.}
  \label{fig:intro_performance}
\end{figure}

\stitle{Motivations and challenges} Scalable vector search requires distributed execution across machines to address both single-machine memory \textit{capacity} and \textit{bandwidth} limitations, especially as billion-scale datasets and high-dimensional vectors (hundreds of dimensions) become common-place today.
More critically, vector search is usually \textit{memory bound} since modern CPUs outpace memory bandwidth and each query accesses numerous vectors~\cite{reorder}. As shown in Figure~\ref{fig:intro_performance},  scaling from 8 to 128 threads on a single machine improves query throughput (QPS)  by only 3-4$\times$, far below the ideal 16$\times$. This is because memory bandwidth (204 GB/s for our machine) is saturated and thus increasing the number of threads beyond a point (48 in our experiments) will not improve QPS. For the same reason, disk-based systems~\cite{diskann,spann}, which keep vectors on SSDs, are also limited by disk bandwidth. In comparison, distributed solutions can leverage the aggregated memory bandwidth of multiple servers for higher throughput.

However, distributed vector search faces a critical tension between \textit{computation efficiency} and \textit{communication efficiency}. The simplest solution is \textit{independent sharding} (called \textit{Shard}), as shown in the middle of Figure~\ref{fig:intro:app}. Shard assigns vectors across machines, each constructing a local index on its vectors. 
A query is first broadcast to all machines for local processing, and then the local search results are merged to obtain the global top-$k$ most similar vectors.
While simple and adopted by popular vector databases such as Milvus~\cite{milvus} and Weaviate~\cite{Dilocker_Weaviate}, \textit{Shard} underperforms: 
Figure~\ref{fig:intro_performance} shows only 4-6$\times$ QPS gain on 16 machines. This is because the query complexity of proximity graph is \textit{sublinear} w.r.t. dataset size (i.e., $\log N$), and thus partitioning a large dataset into $M$ (i.e., the number of machines)  independent sub-datasets means that more similarity computations are used to process each query compared to organizing all vectors in a single index (i.e., $M\log (N/M)> \log N$, where $N\gg M$). 

To fully exploit the pruning power of vector indexes and improve computational efficiency, a natural solution is to utilize a global index on the entire dataset. As shown in the rightmost plot of Figure~\ref{fig:intro:app}, \textit{Global} keeps some vectors and their adjacency lists on each machine, but the vectors on one machine may connect to the vectors on other machines. Each query is assigned to a machine for processing, and when the graph traversal visits a node, some of its neighbors may reside on remote machines. In this case, network communication is required to fetch these neighbor vectors for computation. However, as shown in Figure~\ref{fig:intro_performance}, the QPS of \textit{Global} is also far below ideal linear scaling. This is because \textit{Global} communicates many vectors over the network, which are large in size and easily saturate network bandwidth. Moreover, although the latency of one-sided remote direct memory access (RDMA) communication (at 2-3 $\mu$s) is much shorter than TCP/IP~\cite{fent2020low,kalia2014using}, it is still over 10$\times$ of local memory access. Moreover, since a query needs many communication operations, and the next node to visit can only be determined after updating the candidate queue with the remote vectors, we observe the query latency of \textit{Global} to be 10-20$\times$ of a single machine.

\stitle{Our solution: CoTra} We present CoTra, a distributed vector search system that resolves the tension between computation efficiency and communication efficiency. 
As shown in Figure~\ref{fig:intro_performance}, the QPS of \sys is significantly higher than both \textit{Shard} and \textit{Global} and is over 0.8$\times$ of ideal linear scaling. To our knowledge, CoTra is the first to run a holistic proximity graph index over multiple machines and achieve close to linear scaling w.r.t. single-machine proximity graph.

\begin{figure}[!t]
  \centering
  \includegraphics[width=0.8\linewidth]{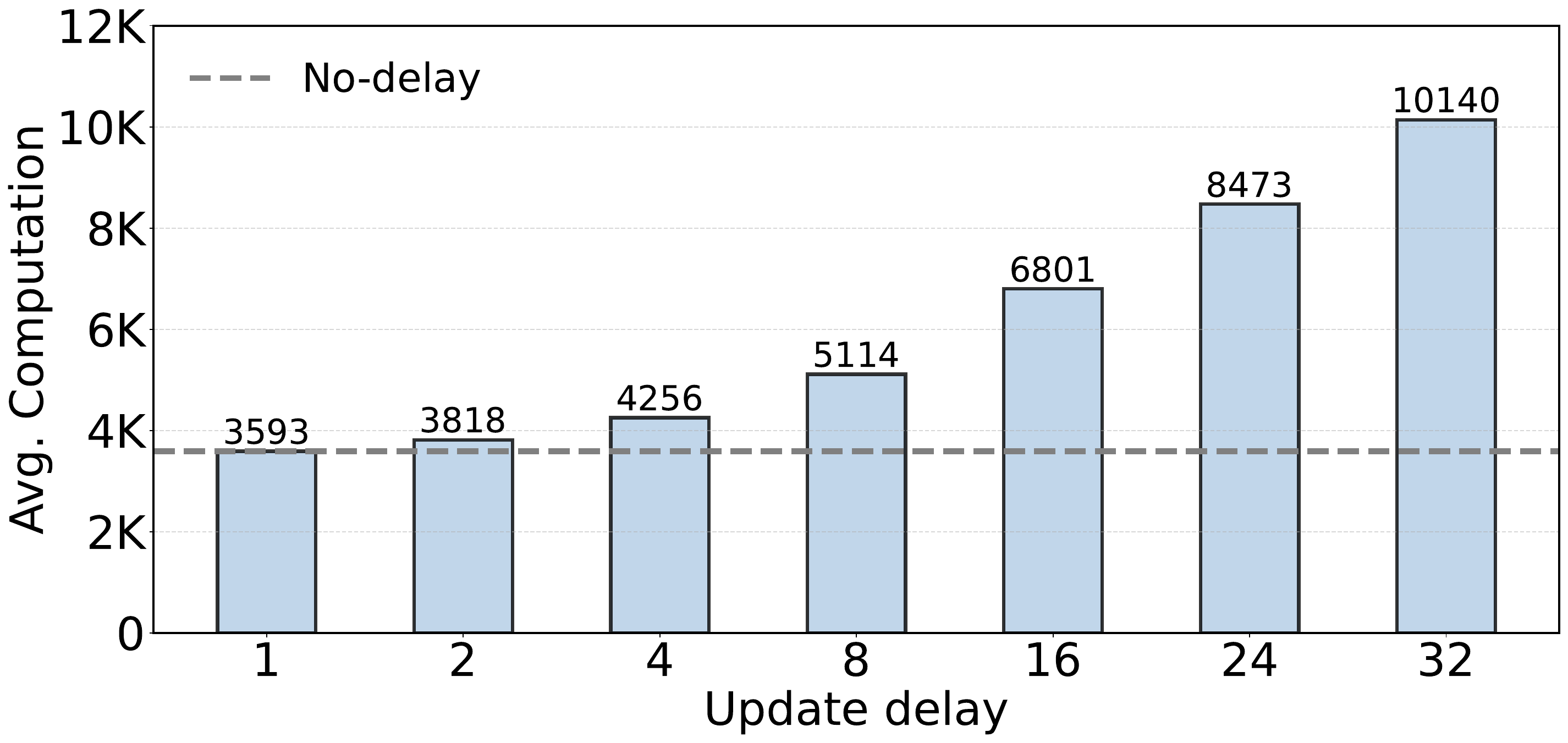}
  \caption{Impact of \textit{candidate queue update delay} on similarity computations (SIFT100M dataset, 0.95 recall).}
  \label{fig:delay_profiling}
\end{figure}

To avoid the network saturation of \textit{Global}, we start with \textit{computation-push}, which pushes the tasks of computing distances to remote machines instead of pulling large vectors. Although computation-push effectively reduces network traffic, its QPS remains low. This is because the remote distance computation tasks have unstable delays, which degrades the candidate queue that is crucial for proximity graph traversal. In particular, if a node $u$ has smaller distance than the current best candidate $v$ but is returned late, the graph traversal will visit $v$, which is suboptimal and thus wastes computation. This is  illustrated in Figure~\ref{fig:delay_profiling}, where we delay the updates to the candidate queue after visiting some nodes and computation escalates with moderate delays (e.g., 16). We also tried several data partitioning techniques but found it difficult to concentrate the vectors required by each query to a single machine to avoid communication.    

The failure of \textit{computation-push} suggests that candidate quality should be maintained when reducing communication, and \sys achieves this goal with algorithm-system co-designs. In particular, K-means is utilized to partition vectors to machines in a similarity-based manner. For each query, this allows to distinguish  machines into \textit{primaries} that host many accessed vectors and  \textit{secondaries} that host a few accessed vectors. Moreover, vectors on the primary machines are generally closer to the query. As such, we use different execution strategies for the primary and secondary machines. The primary machines run \textit{Co-Search} and work like single-machine search but their local candidate queues are synchronized at regular intervals to ensure the quality of the candidates. The secondary machines run \textit{Push-Pull} by sending vectors or computing distances for the primary machines on demand; they may have long delays to update the candidate queue but the degradation is limited since their vectors are usually far from the query.

\sys also incorporates a suite of system optimizations to improve efficiency. First, as the basic operations, both \textit{Co-Search} and \textit{Push-Pull} are implemented efficiently over RDMA. Second, task scheduling is conducted on each machine to overlap computation and communication and avoid long synchronization delays for the candidate queues. Third, communication tasks on the same machine are carefully batched to reduce the required IO operations. Finally, we tailor the distributed storage format of the proximity graph index for efficient communication and task execution. We also implement a distributed procedure to build the global proximity graph index for large vector datasets that do not fit in a single machine.             

We evaluate \sys on four large-scale vector datasets with up to 1 billion vectors, comparing it against \textit{Shard}, \textit{Global}, and a state-of-the-art distributed vector database. The results show that \sys consistently achieves higher QPS than all baselines and has close-to-linear scaling for QPS when increasing the number of machines. Our analyses and ablation studies also validate the effectiveness of our designs.

To summarize, we make the following main contributions.

\squishlist
\item We motivate distributed vector search by scaling up both memory capacity and bandwidth. We also identify the key challenge for good scalability as the tension between computation and communication efficiency with profiling.

\item We propose a collaborative search mechanism that enables fine-grained coordination across the machines for vector search, which maintains  computation efficiency at a low communication cost.

\item We tailor a suite of system optimizations for \sys, which includes task scheduling, communication batching, index storage, and distributed index building.
\item We open-source \sys and extensively evaluate its performance, providing insights into the design choices.
\squishend

\section{Background}

In this part, we introduce the basics of vector search and RDMA to facilitate the subsequent discussions.  

\subsection{Vector Search and Proximity Graph Index}

Vector search is also known as nearest neighbor search (NNS) and defined as follows. 
\begin{definition}
Given a vector dataset $\mathcal{X} = \{x_1, x_2, \cdots, x_N\} \subset \mathbb{R}^d$ with size $N$, a query vector $q \in \mathbb{R}^d$, and the number of required vectors $k$, return a set of vectors $\mathcal{S}\subset\mathcal{X}$ that satisfy
\[\Vert x_i-q\Vert \le \Vert x_j-q\Vert, \forall x_i \in \mathcal{S},x_j\in \mathcal{X}-\mathcal{S} \ \text{and} \ |\mathcal{S}| = k.  \]
\end{definition}
Besides Euclidean distance, the similarity function may also be inner product or cosine similarity. Since embedding vectors usually have a high dimension (e.g., 100 or even $>$1000), exact NNS requires a linear scan. 
Thus, approximate NNS (ANNS) is often employed in practice, which returns most rather than all of the top$k$ neighbors to trade for efficiency. The result quality of ANNS is measured by \textit{recall}, which is defined as $|\mathcal{S}\cap \mathcal{S}'|/k$, where $\mathcal{S}'$ is an approximate result set with $k$ vectors. Applications typically require a high recall (e.g., 0.9 or 0.95). ANNS needs to handle large vector datasets (e.g., for RAG with trillions of tokens~\cite{shao2024scaling}) and provide high query processing throughput (QPS) (e.g., for search and recommendation~\cite{spann,covington2016deep}). A short query latency (e.g., 10-100ms) is also required for the quality-of-service (QoS) of  user-facing applications.

\begin{figure}[!t]
  \centering
  \includegraphics[width=0.7\linewidth]{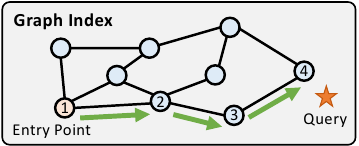}
  \caption{Vector search with proximity graph index.}
  \label{fig:bg_anns}
\end{figure}


\begin{algorithm}[!t]
\caption{Graph Traversal for Vector Search}
\begin{algorithmic}[1]
\State \textbf{Input:} Graph $G$, query $q$, result count $k$, queue size $L$
\State \textbf{Output:} $k$ similar vectors to $q$
\State Initialize a size-$L$ priority queue $Q$ with $(v_0, \Vert v_0-q\Vert)$
\While{$Q$ has unvisited node}
    \State Read the most similar but unvisited node $v$ in $Q$
    \For{each neighbor $u$ of $v$ in $G$}
        \If{distance $\Vert q-u \Vert$ is not computed}
        \State Compute $\Vert q-u \Vert$
        \State Try to insert $(u, \Vert q-u \Vert)$ into $Q$ 
        \EndIf
    \EndFor
\EndWhile
\State \Return The $k$ vectors with the smallest distances in $Q$ 
\end{algorithmic}
\label{alg:beam-search}
\end{algorithm}

Proximity graph~\cite{approximate_graph} is the state-of-the-art index for vector search~\cite{approximate_graph} and requires much fewer distance computations to reach the same recall than other indexes (e.g., IVF and LSH). The key idea is to connect each vector with its similar vectors to form a graph, as illustrated in Figure~\ref{fig:bg_anns}. Vector search is conducted by the graph traversal procedure in Algorithm~\ref{alg:beam-search}. $Q$ is a min priority queue and called the \textit{candidate queue}. Line 3 of Algorithm~\ref{alg:beam-search} initializes $Q$ with the entry node (i.e., $v_0$); lines 4-9 checks the most similar but unvisited node $v$ in $Q$ by computing distances for $v$'s neighbors. Note that the visited nodes can still reside in $Q$ (i.e., line 5 uses $\mathsf{read}$ instead of $\mathsf{pop}$, and $Q$ meets the size constraint $L$ by discarding the vectors with large distances. Users configure $L$ to control the quality of search results, with a larger $L$ yielding higher recall at the cost of longer processing time. Proximity graph has good performance because (i) the graph identifies good candidates, i.e., if a vector is similar to the query, its neighbors are also likely to be similar, and (ii) by visiting the most similar vector each time, the graph traversal gradually move to the results of vector search, as shown in Figure~\ref{fig:bg_anns},

\subsection{Remote Direct Memory Access}

Remote Direct Memory Access (RDMA) is a networking technology for high-throughput, low-latency communication between servers. Different from TCP/IP-based networks, which incur significant overhead from kernel processing and data copying, RDMA allows to bypass remote operating systems and directly read from or write to remote memory. 
Modern RDMA networks achieve bandwidths of 40-400 Gbps~\cite{fasst,rdma2}, far exceeding the 10-25 Gbps of Ethernet, and the end-to-end latencies are in microseconds (i.e., us). RDMA has been widely employed to design efficient distributed systems such as databases~\cite{rdma1} and machine learning~\cite{metascale}. 

RDMA operations are abstracted as \emph{verbs}, a set of low-level primitives including \textit{one-sided} verbs and \textit{two-sided} verbs. One-sided verbs (e.g., \texttt{READ}, \texttt{WRITE}, and atomic operations like \texttt{compare-and-swap/CAS}) execute without remote CPU involvement and thus are ideal for latency-critical tasks. Two-sided verbs (e.g., \texttt{SEND/RECV}) resemble TCP/IP semantics, i.e., require the coordination between sender and receiver CPUs, and are typically used for control messages or compatibility.

The key challenge of using RDMA for vector search is to balance between communication latency and bandwidth utilization. In particular, one-sided RDMA verbs have low latency (e.g., 2-3$\mu$s in our cluster when the payload is below 2KB) but can only access contiguous remote memory for read/write operations. However, vector search follows the proximity graph to conduct fine-grained random accesses to the vectors. Moreover, the throughput of RDMA communication is constrained by both network bandwidth and IO operations per second (IOPS) due to the physical limits of NICs~\cite{wukong}. For instance, to fully utilize the 56 Gbps bandwidth in our cluster, the payload must exceed 1KB due to limited IOPS. However, an individual vector may be smaller than 1KB, e.g., a 128-dimension float vector takes only 128B.

\section{Distributed Collaborative Traversal}

We build \sys as a system that unleashes the power of distributed vector search over multiple machines by efficiently scaling up both memory capability and bandwidth based on RDMA. It aims to run distributed search on the partitioned global graph with good scalability, i.e., the query processing throughput (QPS) should increase linearly with the number of machines. For this purpose, we design a collaborative graph traversal procedure to run the proximity graph search algorithm  (i.e., Algorithm~\ref{alg:beam-search}) for each query over multiple machines with high efficiency.

\stitle{Data layout} To fully utilize the pruning power of vector index, \sys adopts a holistic proximity graph over the entire dataset, and we will discuss how to build the graph in Section~\ref{sec:design}. As shown in Figure~\ref{fig:intro:app}(c), each machine keeps a similar number of vectors and the adjacency lists of these vectors in the graph index. Since the graph is holistic, the vectors on one machine may connect to the vectors on other machines. In particular, we partition the  vector dataset using balanced K-means and assign each partition to one machine. This is intended to make the vectors on one machine similar to each other and concentrate the vectors accessed by each query to a small number of machines for good locality. 

\subsection{Observations and Design Insights}
\label{sec:sys_setting}

\begin{figure}[!t]
  \centering
          \includegraphics[width=0.85\linewidth]{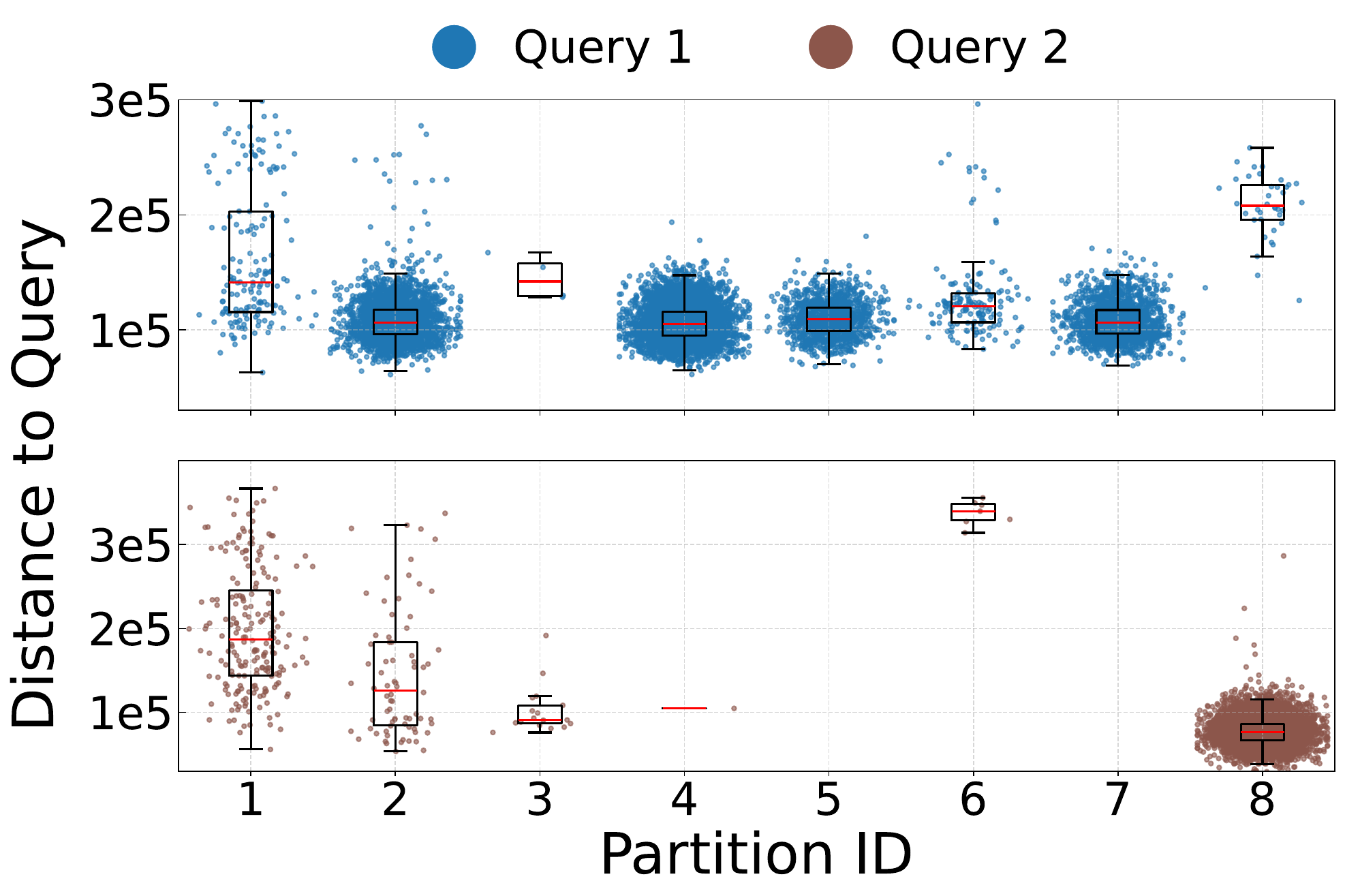}
        \caption{Access patterns of two sample queries. The dataset is SIFT100M and clustered into 8 partitions. Point clouds denote the number of accessed vectors in each partition, while box plots show the distances of these vectors to query.}
        \label{fig:obs:primary_partition}
\end{figure}

Before introducing our collaborative traversal, we present the insights that motivate its designs.

\stitle{Data locality} We first check how effective K-means partitioning is in improving access locality. If each query mostly accesses the vectors on one machine, and we assign the query to this machine for processing, communication will be small. Figure~\ref{fig:obs:primary_partition} shows that this is not the case, i.e., although Query 2 mostly accesses the vectors on Partition 8, the vectors accessed by Query 1 are scattered  more uniformly over the partitions. By averaging over many queries, we find that 73.8\% of the accessed vectors reside on the hottest partition, and note that the hottest partitions are different for queries. As such, if we assign each query to its hottest partition for processing, about 25\% of the vector accesses need to go over the network. This will still make network the bottleneck due to the large bandwidth disparity between memory and network. For instance, our machine has a memory and network bandwidth of 204 GBps and 56 Gbps, respectively; to keep up with local in-memory processing and read  25\% of the vectors over network, the required bandwidth should be 350 Gbps. As such, beside K-means partitioning, other designs are still required to reduce communication costs.

We have also tried more sophisticated data partitioning. That is, we first build the proximity graph and then use a min-cut to partition its nodes over the machines by minimizing the number of cross-partition edges. However, access locality only improves marginally over K-means. This is because some queries inevitably reside on the boundaries of partitions. As such, we choose K-means due to its simplicity. 

In Figure~\ref{fig:obs:primary_partition}, for each query, we can clearly distinguish between hot partitions with many accesses (e.g., Partitions 2, 4, 5, 7 for Query 1, and Partition 8 for Query 2) and cold partitions with a few accesses (e.g., Partitions 1, 3, 6, 8 for Query 1). We call the hot ones \textit{primary partitions} and the cold ones \textit{secondary partitions}.

\vspace{0.3em}
\textbf{Insight 1.} \textit{K-means partitioning improves data locality and allows to distinguish primary and secondary partitions for each query. However, communication is still the bottleneck.}

\stitle{Candidate quality} As vector search is approximate in nature, we do not really need to faithfully execute Algorithm~\ref{alg:beam-search}. To reduce the communication cost, we can flexibly control the amount and frequency of the communications. However, these communication designs should limit their impacts on computation efficiency, i.e., the number of distance computations required by each query to reach the target recall. In particular, we find that the computation efficiency of Algorithm~\ref{alg:beam-search} is decided by the candidate queue $Q$. That is, we may skip or delay some distance computations for the vectors on remote machines; if these computations involves a node (i.e., vector) $u$ that is better than the current best unvisited node $v$ in $Q$, i.e., $\Vert q-u \Vert<\Vert q-v \Vert$, visiting $v$ may lead the graph traversal to detours and wastes computation. Figure~\ref{fig:delay_profiling} also shows that computation increases if the candidate quality is degraded by update delays. Conversely, if the candidate queue can produce the same node $u$ to visit as single-machine execution, computation efficiency will not be affected.  

The above analysis suggests that to avoid degrading the quality of the candidate queue, our communication designs should prioritize high-quality vectors (i.e., those with small distances to the query) and discriminate the low-quality vectors (i.e., those with large distances). Figure~\ref{fig:obs:primary_partition} shows that the primary and secondary partitions are informative about candidate quality because vectors from the primary partitions (e.g., Partitions 2, 4, 5, 7 for Query 1) generally have smaller distances to the query than vectors from the secondary partitions (e.g., Partitions 1, 3, 6, 8 for Query 1).

\textbf{Insight 2.} \textit{
Vectors with small distances to the query should prioritized to keep the quality of the candidate queue. Vectors from the primary partitions are generally closer to the query.}

\begin{figure}[!t]
  \centering
  \includegraphics[width=0.99\linewidth]{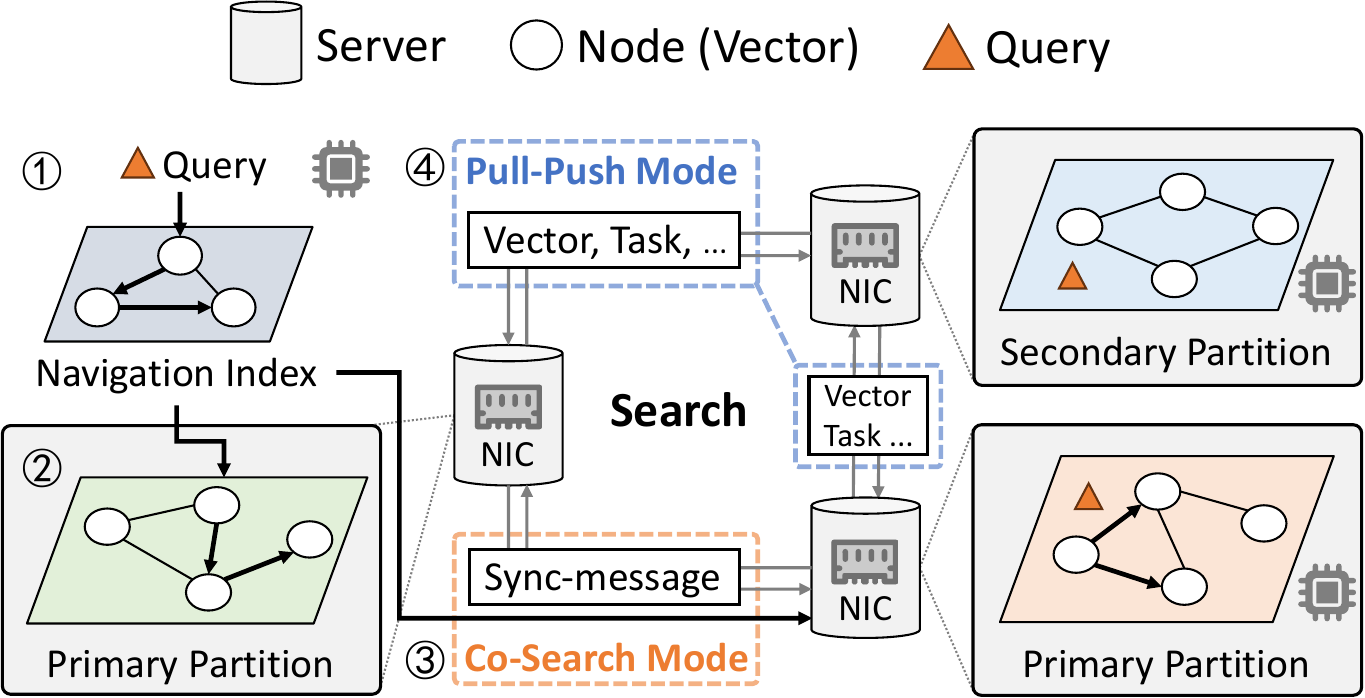}
  \caption{The execution workflow of \sys.}
  \Description{}
  \label{overview}
\end{figure}

\subsection{Collaborative Graph Traversal Overview}

Our observations in Section~\ref{sec:sys_setting} suggest that for each query, the primary partitions with many accessed vectors at close distances and secondary partitions with a few accessed vectors at far distances should be treated differently. Therefore, as shown in Figure~\ref{overview}, we first use a \textit{navigation index} to efficiently distinguish the primary and secondary partitions. Then, we use different modes to coordinate the machines (i.e., partitions) for collaborative graph traversal. In particular, the primary partitions work in \textit{Co-Search} mode by maintaining their own candidate queues, running distance computations on local vectors, and sending distance computation requests to the other partitions (i.e., both primary and secondary). The local candidate queues of the primary partitions are synchronized at regular intervals to ensure \textit{bounded delay} and thus quality of the candidates. In contrast, the secondary partitions work in \textit{Pull-Push} mode by sending vectors or computing distances as requested by the primary partitions without maintaining local candidate queues on them. 

The key idea of our design to \textbf{maintain candidate quality at low communication costs}. First, the primary partitions do not exchange vectors and distances because many vectors are accessed on them; instead, the local candidate queues are much smaller for exchange. Second, to ensure candidate quality, the primary partitions are synchronized for bounded delay because their vectors are generally close to the query; tasks on the the secondary partitions have looser synchronization requirements because their vectors are usually far from the query. Third, candidate queues are not maintained on the secondary partitions because a few vectors are accessed on them, and thus the communication costs for exchanging vectors and distances are moderate. 

\begin{figure*}[!t]
  \centering
  \includegraphics[width=0.95\linewidth]{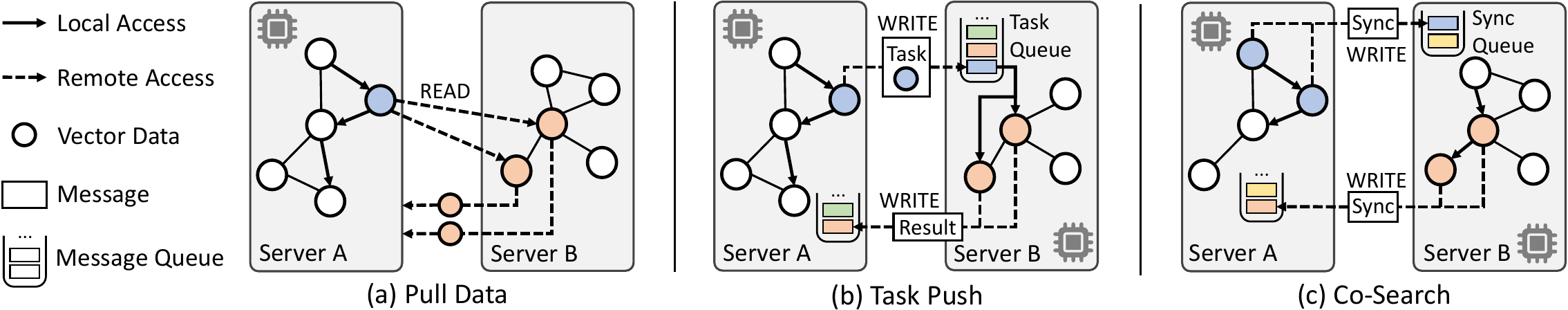}
  \caption{\sys communication operations. (a-b): When visiting the blue node, its orange neighbors are required on remote machines; (c): Co-Search conducts parallel and synchronizes query states between servers A and B.}
  \Description{}
  \label{fig:exe_model}
\end{figure*}

\label{sec:collab_overview}
\stitle{Navigation Index} We sample a small portion (e.g., 1\%) of vectors from the dataset and build a proximity graph index on these vectors as the navigation index. The navigation index is replicated over all machines, and each query is first assigned to a random machine for initialization. The machine searches the query with the navigation index to determine its primary and secondary partitions. In particular, we find the top-$k$ neighbors of the query in the navigation index and count how many of these neighbors reside on each partition (of the entire dataset). A partition is primary if it holds more than $k/M$ of these neighbors ($M$ is the number of partitions) and secondary otherwise.

To speed up query processing, we use the top-$k$ neighbors from the navigation index to initialize the local candidate queues of the primary partitions. Each primary partitions collect the neighbors that reside on them, while the neighbors on the secondary partitions are sent to the primary partition that hosts the largest number of these top-$k$ neighbors.

\begin{algorithm}[!t]
\caption{Collaborative Traversal Search}
\label{alg:hybrid_search}
\begin{algorithmic}[1]
\State \textbf{Input:} Graph $G$, query $Q$, number of result vectors $k$
\State \textbf{Output:} $k$ vectors that are similar to $Q$
\State $\mathsf{cand\_queue} \gets \text{NavigationIndex}(Q)$
\State $\text{DispatchSubQuery}(\mathsf{cand\_queue})$

\While{Changed($\mathsf{cand\_queue}$) or !Terminate($Q$) }
    
    \State $\mathsf{Nodes} \gets \mathsf{cand\_queue}.\text{top\_unvisited}()$
    \State $\mathsf{Machines}, \mathsf{Tasks}, \mathsf{SyncMsg} \gets G.\text{getNgh}(\mathsf{Nodes})$

    \For {$\mathsf{machine} \in \mathsf{Machines}$}
        \If {$\mathsf{machine}$ is primary}
            \State $\text{PostSync}(\mathsf{SyncMsg})$ // Co-Search mode
        \Else
            \State $\text{PullPush}(\mathsf{Tasks})$ // Pull-Push mode
        \EndIf
    \EndFor
    \State $\mathsf{SyncMsg} \gets \text{LocalTraversal}(\mathsf{Tasks})$
    \For {$\mathsf{Result}, \mathsf{RecvSync} \gets \text{PullMsg}()$}
        \State $\mathsf{cand\_queue}$.update($\mathsf{Result}$, $\mathsf{RecvSync}$)
    \EndFor 
        \State $\text{TerminationDetection}(Q)$
\EndWhile
\State \Return{$\mathsf{cand\_queue}$.top($k$)}
\end{algorithmic}
\end{algorithm}

\stitle{Collaborative traversal search} Algorithm~\ref{alg:hybrid_search} details query processing with our collaborative traversal. 

In lines 3-4, the navigation index is searched to identify primary and secondary machines and dispatch sub-queries (i.e., candidate queues) to the primary machines (Figure~\ref{overview} \textcircled{1}$\xrightarrow{}$\textcircled{2}), which start graph traversal in lines 5-12. Local traversal follows Algorithm~\ref{alg:beam-search}, i.e., each iteration extracts the most promising candidate from the candidate queue, traverses its neighbors, and packs the computation tasks into different message types, such as computation tasks and synchronization messages, based on the machines where their targeted vectors reside. 

Each sub-query processes its local tasks while concurrently updating the synchronization messages (line 13). Based on the messages received from the query message queues, which contain both remote computation results and synchronization data from other primary machines, a sub-query updates its candidate queue (lines 14-15). A termination detection algorithm (line 16), which will be detailed in Section~\ref{subsec:implementation}, activates when the candidate queue is empty.

\subsection{Execution Mode Details}
\label{sec:execution-model}

We implement the Pull-Push and Co-Search modes for collaborative traversal efficiently over RDMA. Figure~\ref{fig:exe_model} provides an illustration, and we present the details as follows.

\stitle{Pull-Push mode} We use it pull vectors from and push distance computation tasks to the secondary machines.
\squishlist
\item \textit{Pull-Data}:
As shown in Figure~\ref{fig:exe_model}(a), when visiting a candidate, some of its neighbors may reside on a remote machine. \sys can determine the remote memory addressees of these target vectors according to their IDs and retrieves them via RDMA one-sided primitives. A pre-allocated local buffer is used to store the retrieved vectors, and an asynchronous one-sided READ primitive is adopted to avoid communication stall.

\item \textit{Task-Push}: To reduce the bandwidth consumption for transferring large vectors, \sys can use task-push as shown in Figure~\ref{fig:exe_model}(b). Specifically, the IDs of the vectors to be computed are written into the remote machine's task queue using one-sided RDMA writes. The remote machine retrieves tasks from the queue, 
performs computations on the specified vectors, and returns only the distances to the requesting machines. Note that each query only needs to be sent to the machines once.
\squishend

We use \textit{Pull-Data} when a remote machine has fewer than or equal to two neighbors of the currently visited node and \textit{Task-Push} otherwise. This is because \textit{Pull-Data} does not involve the CPUs of remote machines and thus has shorter latency. This is favorable but it can incur excessive bandwidth consumption when there are many remote neighbors. Empirically, we observe that setting the threshold as 2 balances between latency and bandwidth.

\stitle{Co-Search mode} The primary machines use the Co-Search mode to exchange query states as shown in Figure ~\ref{fig:exe_model}(c). In particular, the exchanged messages include
\squishlist
\item new candidates that are inserted into the candidate queue since the last synchronization. 

\item the current distance upper bound for a node to be added to the candidate queue. A node must have smaller distance to query than the bound to enter the candidate queue.

\item vectors whose distances need to be computed on the remote machines since the last synchronization.
\squishend
These messages grouped by their target machines and packed together for sending. Co-Search conducts synchronization after visiting a fixed number of candidates (set as 4 by default) to ensure good quality for the candidates. Using the synchronization messages, each primary machine updates its local candidate queue and distance upper bound (by taking the minimum of all bounds), and checks if the candidates to be traversed locally have been visited by the other machines. If not, it traverses the candidates to update the local candidate queue as in Algorithm~\ref{alg:beam-search}. The distance upper bound is used to reject candidates that cannot be added to candidate queue.

\begin{figure}[!t]
  \centering
  \includegraphics[width=0.90\linewidth]{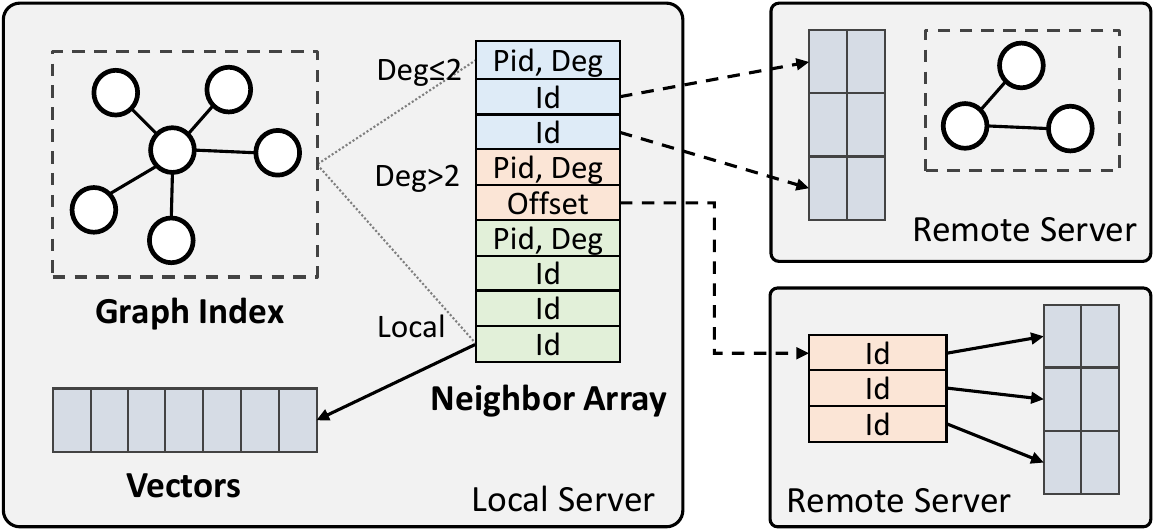}
  \caption{Optimized graph layout.}
  \label{graph_layout}
\end{figure}

\section{Optimizations and Implementations}\label{sec:design}

\subsection{RDMA-friendly Graph Layout}
To reduce communication costs, we tailor a graph storage format for vector search and RDMA. In particular, we store the proximity graph index in a decoupled manner, where the neighbors of a node on a remote machine are stored on the remote machine. As shown in Figure~\ref{graph_layout}, the adjacency list of each node consists of local edges and pointers to remote edges. To work with the hybrid \textit{Pull-Push} mode, each node stores pointers for high-degree remote neighbor array and IDs for low-degree remote neighbors. Using this storage format, the \textit{Pull-Push} mode can further reduce network bandwidth, as the transmission of large edges is replaced by the pointers. 

To conduct efficient memory access, we implement fine-grained prefetching. Vector search on proximity graph conducts many random memory accesses, including local vector reads, graph adjacency list traversals, and visit list operations. We proactively issue prefetch instructions to overlap data transfers with computation. When constructing the graph index, we pre-group each node's neighbors by their respective remote machines and assign metadata to neighbors that reside on the same machine. The metadata includes both the ID of the remote partition and the number of neighbors in that remote partition. Moreover, we employ fixed-size neighbor arrays for inline storage allocation, enabling direct offset calculation from node IDs. This eliminates the initial array access cache misses through predictable memory addressing.

\subsection{Task Scheduling for High CPU Utilization}

To improve throughput and pipeline computation and communication, \sys processes multiple queries simultaneously. However, this may lead to increased query latency as some queries can be starved for computation. 
To balance query latency and throughput, \sys uses C++ coroutines to break down each query into fine-grained operators, each representing a single hop. By scheduling these operators, \sys avoids starvation and achieves inter-query load balancing.

\begin{figure}[!t]
  \centering
  \includegraphics[width=0.90\linewidth]{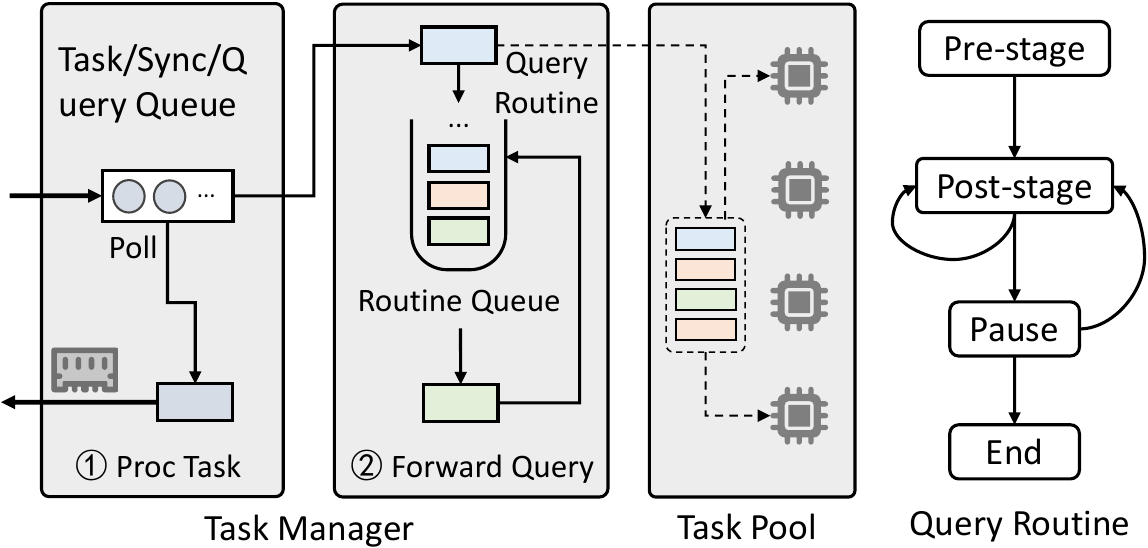}
  \caption{Query states and task scheduling.}
  \label{fig:query_manage}
\end{figure}

\stitle{Query routine}
\sys creates a query routine for each query, which is a state machine. As shown in Figure~\ref{fig:query_manage}, each query transits through four states from start to finish:

\squishlist
\item \textit{Pre-Stage}: As described in Section~\ref{sec:collab_overview}, in this phase, the query searches the navigation index to determine its primary and secondary machines. Sub-queries are then dispatched to the primary machines. Each primary machine creates a query routine and enters the second state.

\item \textit{Post-Stage}: In this phase, the query performs local search and traversal, computes distances for candidate nodes, updates the candidate queue, and synchronizes messages with other primary machines. After processing a fixed number of nodes in the current candidate queue, it transits to the next state.

\item \textit{Pause}: In this phase, the query enters a suspending state and initiates the distributed termination algorithm. If new synchronization messages received in this period can provide new candidate nodes, the query is reactivated and returns to the post-stage. Otherwise, it waits for the termination token using the termination detection algorithm.

\item \textit{End}: When a query successfully detects termination, its local search results are sent back to the originating machine, which merges these results for final top-$k$ vectors.

\squishend

\noindent As shown in Figure~\ref{fig:query_manage}, we manage queries using a routine queue. Each thread repeatedly performs two procedures, i.e., process remote tasks and advance query routines. When processing remote tasks, as described in Section~\ref{sec:execution-model}, the thread retrieves computation tasks from the task queue and returns the results. After processing the task queue, the thread pops a query routine from the routine queue and executes one step. If the query routine remains in the post-stage state, it is re-inserted into the routine queue for the next round. If it transits to the Pause or End state, it is not re-inserted into the routine queue.

\stitle{Load balancing} Due to the dynamic nature of queries, the number of sub-queries handled by each thread may become uneven. To achieve load balancing, we employ a multi-threaded task pool to manage query computations. The tasks created by query routines are placed in a queue shared by all threads. After processing the task queue, the threads check if the query routine has sub-queries and perform task stealing from the shared queue. 

\subsection{Implementation Details}\label{subsec:implementation}
We implemented \sys with 27K lines of C++ code, including distributed query processing and index building. \sys can be deployed on general RDMA clusters and supports different graph indexes.

\stitle{Communication batching and pipeline}
To efficiently utilize RDMA, we merge small messages (e.g., distances of vectors to the query). By reducing the number of messages, the IOPS bottleneck of NICs is tackled. To support efficient message passing and avoid excessive memory consumption, we design a one-sided write queue  based on the circular buffer queue. When a machine receives a certain amount one-sided write operations, it returns the IDs of the released buffers, allowing the remote message buffers to re-enter the circular buffer queue for reuse.

\stitle{Distributed query termination}
In the \textit{Co-Search} mode, query termination is challenging. Due to the dynamic nature of query processing, a sub-query may still be activated and continue execution after receiving update messages from remote machines even after it has terminated locally. Noticing the similarity between ANNS and the Single Source Shortest Path algorithm, we implement a distributed termination detection algorithm based on Dijkstra-style 2-pass ring termination detection~\cite{termination_detection}. In particular, during query execution, each sub-query is assigned an identifier flag, and only one sub-query holds the termination token at any given time. When a sub-query completes its execution, it passes the token to the next sub-query. The sub-query that receives the token determines the token's value based on whether new computations have been performed since the last time it held the token and the state of its flag. This process is repeated by each sub-query, and if the same token circulates twice among the sub-queries, the query can be terminated safely.

\stitle{Distributed index building} We build a proximity graph index for vector datasets that exceed the memory of a single machine following the replica-based partitioning approach of DiskANN~\cite{diskann}. The idea is to partition the dataset via K-means, send each vector to multiple partitions, and merge the local indexes of the partitions such that the graph is connected due to replicated nodes between the partitions. We divide the entire process into three phases i.e., \textit{dispatch}, \textit{build}, and \textit{merge}. In the dispatch phase, each machine collects the vectors destined for every other machine (as determined by K-means on a sample of the dataset) and sends them to the target machines in bulk. 
Each vector is sent to the $S$ closest machines (with $S=2$ by default in DiskANN). Each machine independently constructs a local proximity graph on its assigned vectors. 
In the merge phase, the machines perform a distributed merge operation to de-duplicate the nodes and generate the final graph. CoTra defaults to using Vamana~\cite{diskann} as the graph index due to its good performance. However, CoTra can also construct and search with other proximity graph indexes such as HNSW~\cite{hnsw} and NSG~\cite{nsg}.

\stitle{Using regular network} 
As \sys mostly communicates distances instead of raw vectors, we observe its bandwidth consumption to be low. As such, the main disadvantage of Ethernet over RDMA is longer communication latency. This increases synchronization delays and thus degrades candidate quality. As a result, more computations may be required for each query. We leave testing and adapting \sys to Ethernet for future work.    

\section{Evaluation}
We conduct extensive experiments to compare \sys with state-of-the-art distributed vector search systems, verify the scalability of \sys, and present systematic analyses of performance gains, as well as an ablation study.
\subsection{Experiment Settings}

\begin{table}[!t]
\centering
\caption{The datasets used in the experiments.}
\label{tab:datasets}
\begin{tabular}{ccccc}
\toprule
\textbf{Dataset} & \textbf{Num} & \textbf{Dim} & \textbf{Type} & \textbf{Similarity} \\
\midrule
SIFT & 1B/100M &128 & uint8 & L2 \\
DEEP & 1B/100M & 96 & float32 & L2 \\
Text2Image &100M & 200 & float32 & Inner-product \\
LAION & 100M & 512 & float32 & L2 \\
\bottomrule
\label{dataset}
\end{tabular}
\end{table}

\begin{figure*}[!t]
  \centering
  \includegraphics[width=1\linewidth]{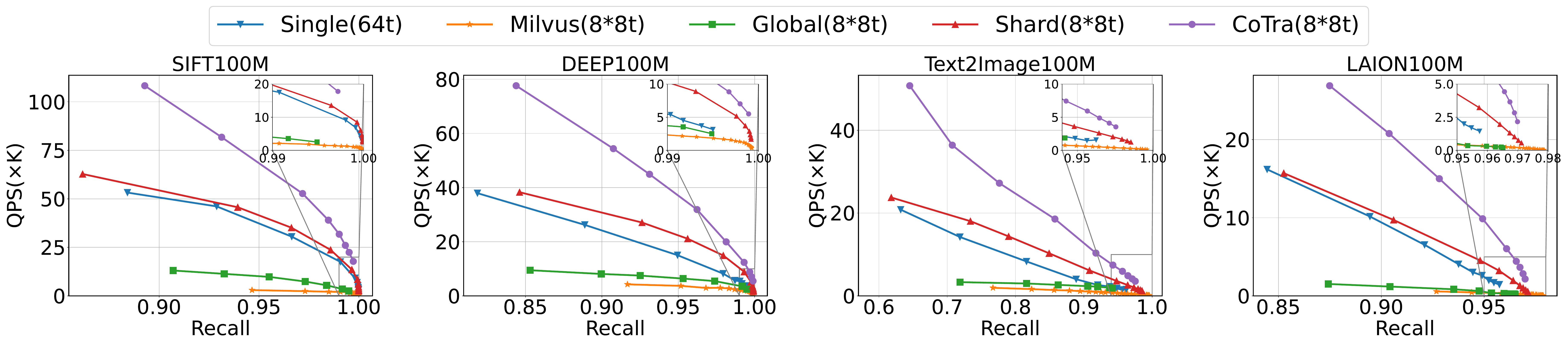}
  \caption{QPS vs. Recall on four datasets with 100 million nodes each. The distributed baseline uses 8 machines with 8 threads per machine, while the in-memory single-machine baseline uses 64 threads.}
  \Description{}
  \label{fig:main_4}
\end{figure*}

\begin{figure*}[!t]
  \centering
  \includegraphics[width=1\linewidth]{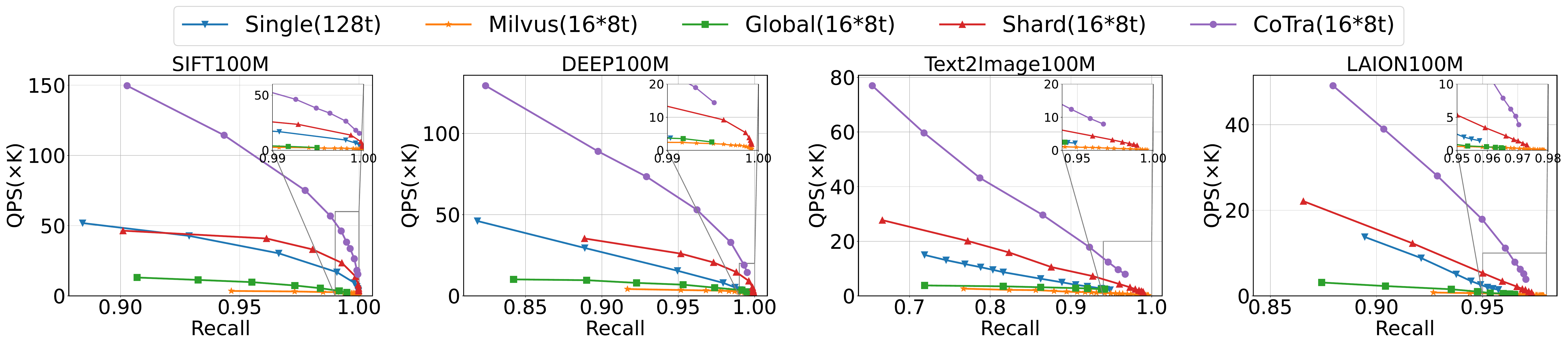}
  \caption{QPS vs. Recall on four datasets with 100 million nodes each. The distributed baseline uses 16 machines with 8 threads per machine, while the in-memory single-machine baseline uses 128 threads.}
  \Description{}
  \label{fig:main_3}
\end{figure*}

\stitle{Datasets}
In our experiments, we evaluate our system on four mainstream benchmark datasets~\cite{simhadri2022results,tellez2024overview}: SIFT~\cite{sift}, DEEP~\cite{babenko2016efficient}, Text-to-Image~\cite {yandextexttoimage}, and LAION~\cite{schuhmann2021laion}. To comprehensively assess performance across different scales, we conduct experiments on the SIFT and DEEP datasets using both 100M and 1B vectors, while 100M subsets are used for evaluations on all four datasets. The data distribution of queries in Text2Image differs from the indexed data, and we verify \sys's capability of handling out-of-distribution data on this dataset. We evaluate our system's search performance on high-dimensional vector datasets by utilizing image embeddings from LAION.

This experimental design enables systematic analysis of system behavior under varying data volumes and different data distribution characteristics. We summarize the dataset configuration in Table~\ref{dataset}.

\stitle{Baselines} We evaluate our system against several representative baseline implementations, including both state-of-the-art single-machine and distributed systems, as well as two RDMA-optimized approaches we developed. The selected baselines include:

\squishlist
    \item \textbf{DiskANN (Single)}\cite{diskann}: State-of-the-art billion-scale graph solution run \underline{fully in-memory} for best performance. We include the pure memory implementation to demonstrate potential memory bottlenecks in single-machine configurations with multiple CPUs and large memory, highlighting the significance of scalability in distributed systems.
    \item \textbf{Milvus}\cite{milvus}: The most popular distributed vector database employing fixed-size partitioned graph indexing, evaluated only on the 100M dataset due to substantial storage requirements (approximately required 2TB for 100M vectors, 20TB for 1B vectors\footnote{\url{https://milvus.io/tools/sizing}}). The fixed-size sharding approach leads to excessive fragmentation when handling large-scale datasets, consequently requiring the search process to traverse numerous shards and incur substantial redundant computations.
    \item \textbf{Global}: RDMA-based implementation using one-sided READ operations for vector data access with K-means partitioning and holistic graph index. This approach is optimized for direct remote memory access via RDMA, eliminating the need for server-side processing during data retrieval operations.
    \item \textbf{Shard}: Fully independent indexing solution operating in scatter-and-gather mode, where queries are scattered across partitions for independent search and results are gathered afterward through RDMA. This design maximizes parallelism in distributed environments while maintaining complete partition independence.
\squishend

The baseline graph index types are Vamana~\cite{diskann}, while HNSW~\cite{hnsw} is used in Milvus. For index construction, we follow the official guidelines and practical recommendations to set the parameters.

\stitle{Testbed}
The experimental evaluation was conducted in two settings. \textcircled{1} The primary testbed comprises a 16-node RDMA cluster, each node equipped with an Intel Xeon Silver 4110 processor (16 cores) and 128 GB of main memory, running on Ubuntu 22.04.4 LTS. The cluster employs 56-Gbps Mellanox MT27500 ConnectX-3 IB RNIC connected to a 56-Gbps InfiniBand switch. \textcircled{2} Another server is a single machine equipped with four Intel Xeon Gold 6252N processors with a total of 768GB of memory, where each NUMA node contains 24 physical cores with 48 logical threads. This server is used to evaluate the in-memory solution, which scales with the number of threads on a single machine.

We use setup \textcircled{2} as our reference point for scalability evaluation, as preliminary testing on smaller-scale datasets confirmed that its performance is comparable to that of individual nodes in setup \textcircled{1}. This validation ensures that meaningful relative performance measurements can be made between the single-machine baseline and distributed configurations. All experimental results are reported as the average of five runs to ensure statistical reliability.

\stitle{Performance metrics}
Following prior works~\cite{nsg,hnsw,spann,diskann,simhadri2022results}, we measure system throughput using queries per second (QPS) that reach a target recall and adopt recall@k (with k=10) as the default accuracy metric, which represents the ratio of search results that are in the ground truth top-k  neighbors. We also observed similar performance trends when setting k = 1 and k = 100 in the micro study.

\subsection{Search Performance and Scalability}

\begin{figure}[!t]
  \centering
  \includegraphics[width=1\linewidth]{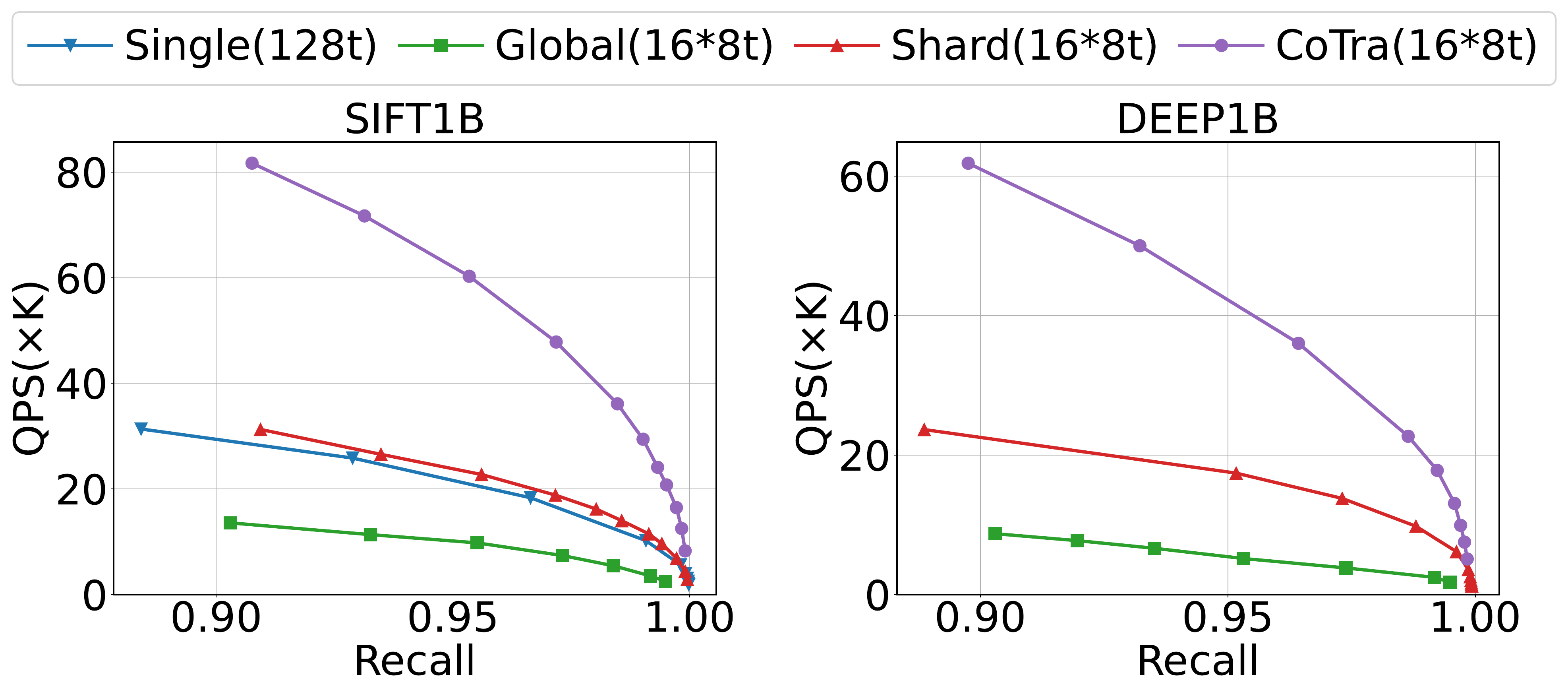}
  \caption{QPS vs. Recall on two 1-billion-node datasets: distributed (16 machines × 8 threads) vs. single-machine (128 threads). The single-machine baseline omits DEEP due to memory constraints.}
  \Description{}
  \label{fig:main_1b}
\end{figure}

\stitle{Performance on 100 million-scale datasets}
\sys demonstrates exceptional scalability and search performance across both 8- and 16-machine configurations. As shown in Figure~\ref{fig:main_4}, \sys achieves a 1.52--2.2$\times$ improvement over the best baseline (\textit{Shard}) on 8 machines, across four datasets (recall@10$\ge$0.95). This performance boost is primarily due to enhanced computational efficiency resulting from the integrity of the graph index and fine-grained execution modes (Section ~\ref{sec:execution-model}).

Figure~\ref{fig:main_3} illustrates the QPS-recall curves for four 100M datasets on 16 machines. The throughput gains are more substantial compared to the 8-machine setup. Specifically, CoTra performs well on the 512-dimensional LAION dataset, outperforming the best baseline by 3.58$\times$ (recall@10 $\ge$ 0.95), and maintains robust performance on the Text2Image dataset, which exhibits distinct recall-QPS characteristics due to out-of-distribution effects, achieving a 2.23$\times$ improvement (recall@10 $\ge$ 0.95). Improvements of 2.12--2.86$\times$ are also observed on the SIFT and DEEP datasets. In contrast, both \textit{Single} and \textit{Global} index configurations fail to scale with increasing computational resources. This is because larger vector sizes exacerbate memory bandwidth bottlenecks in single-machine systems and increase communication overhead in distributed environments.

Across all datasets, the baseline implementation of Shard consistently outperforms Milvus, and CoTra achieves throughput improvements ranging from 8.7$\times$ to 33.3$\times$ compared to Milvus for recall@10$\ge$0.95 on 8- and 16-machine configurations. This is due to Milvus's fixed-shard partitioning strategy, which differs from our machine-count sharding in the Shard index, resulting in more index shards (referred to as segments in Milvus). As the dataset size increases, more shards are introduced, leading to greater computational redundancy and lower throughput in Milvus.

\begin{figure}[!t]
  \centering
  \includegraphics[width=0.9\linewidth]{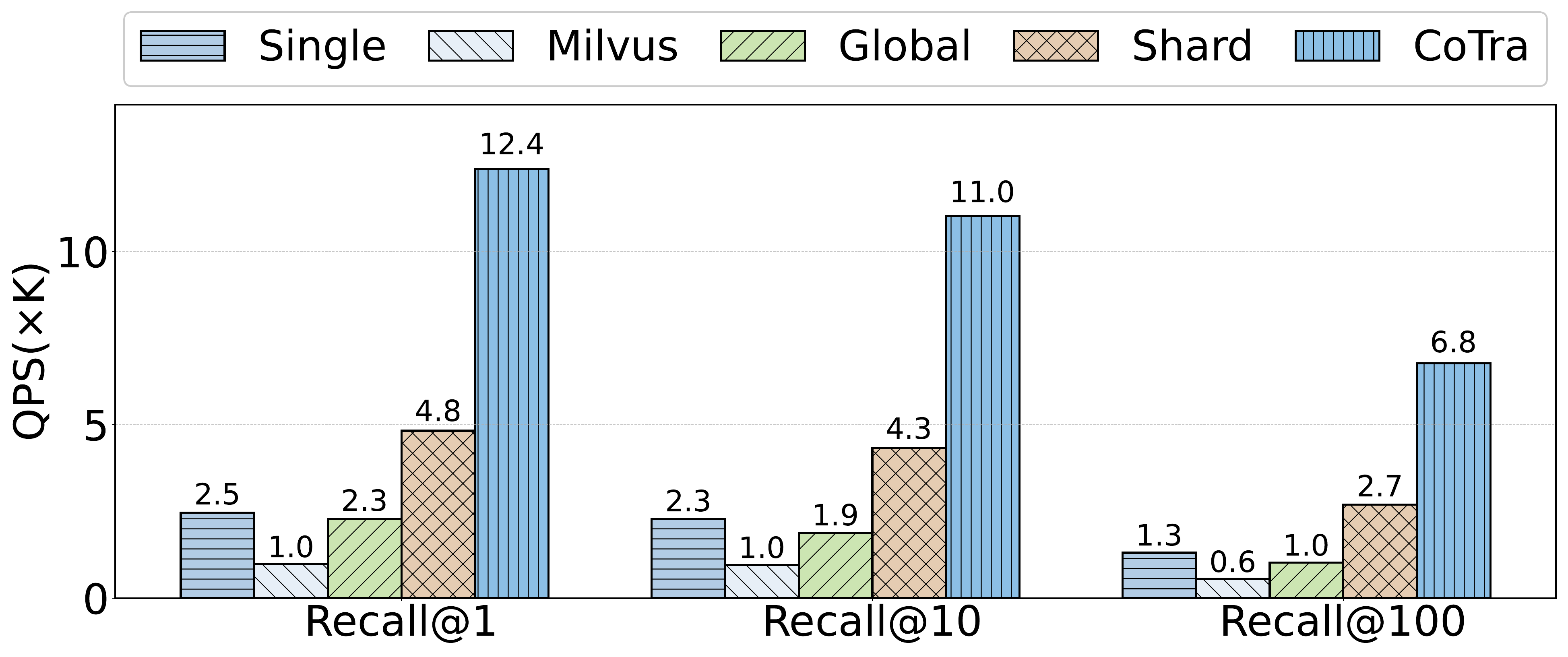}
  \caption{Throughput at 0.95 recall for different top-k on Text2Image100M dataset: distributed baseline (16 machines × 8 threads) vs. single-machine baseline (128 threads).}
  \Description{}
  \label{fig:recall@}
\end{figure}

\begin{table*}[!t]
  \small
  \centering
  \caption{The throughput (QPS) at 0.95 recall and speedup over the single machine (8 threads) for each dataset and solution.}
  \label{tab:new exp-jct-util}
  \setlength{\tabcolsep}{3pt}
  \begin{minipage}{0.48\textwidth}
    \centering
    \scalebox{0.98}{
      \begin{tabular}{lcccc}
        \toprule
        & \multicolumn{4}{c}{\textbf{8 Machine (64 Threads for Single)}} \\
        \cmidrule(r){2-5}
        System & SIFT & DEEP  & Text2Image & LAION\\
        \midrule
        \textbf{Single} & 30.5K (3.4$\times$) & 15.0K (3.7$\times$) &  1.8K (1.5$\times$) & 2.7K (1.7$\times$)\\
        \midrule
        \textbf{Milvus} & 3.0K (0.3$\times$) & 3.7K (0.9$\times$) & 0.7K (0.6$\times$) & 0.4K (0.3$\times$)\\
        \midrule
        \textbf{Global} & 6.3K (0.7$\times$) & 5.9K(1.4$\times$) & 1.4K (1.2$\times$) & 1.0K (0.6$\times$)\\
        \midrule
        \textbf{Shard} & 35.1K (3.9$\times$) & 21.1K (5.1$\times$) &  3.3K (2.9$\times$) & 4.5K (2.8$\times$)\\
        \midrule
        \textbf{CoTra} & 68.9K (7.6$\times$) & 32.1K (7.6$\times$) &  7.2K (6.2$\times$) & 9.9K (6.2$\times$)\\
        \bottomrule
      \end{tabular}
    }
  \end{minipage}
  \hspace{0.15cm}
  \begin{minipage}{0.48\textwidth}
    \centering
    \scalebox{0.98}{
      \begin{tabular}{lcccc}
        \toprule
        & \multicolumn{4}{c}{\textbf{16 Machine (128 Threads for Single)}} \\
        \cmidrule(r){2-5}
        System & SIFT & DEEP & Text2Image & LAION \\
        \midrule
        \textbf{Single} & 30.3K (3.4$\times$) & 15.5K (3.8$\times$) &  2.1K (2.1$\times$) & 3.1K (1.9$\times$)\\
        \midrule
        \textbf{Milvus} & 3.5K (0.4$\times$) & 3.6K (0.9$\times$) & 0.9K (0.9$\times$) & 0.6K (0.4$\times$)\\
        \midrule
        \textbf{Global} & 9.7K (1.1$\times$) & 7.4K (1.8$\times$) & 1.9K (1.9$\times$) & 1.3K (0.8$\times$)\\
        \midrule
        \textbf{Shard} & 40.8K (4.5$\times$) & 26.0K (6.5$\times$) & 4.3K (4.4$\times$) & 5.0K (3.1$\times$)\\
        \midrule
        \textbf{CoTra} & 116.6K (12.7$\times$) & 55.3K (13.4$\times$) & 9.6K (9.8$\times$) & 17.9K (11.2$\times$)\\
        \bottomrule
      \end{tabular}
    }
  \end{minipage}
\end{table*}

\stitle{Performance on 1 billion-scale vector datasets} 
\sys also delivers superior performance on billion-scale vector datasets. Figure \ref{fig:main_1b} presents the query throughput (QPS) versus recall on the SIFT1B and DEEP1B datasets using 16 machines. The performance patterns observed on the SIFT and DEEP datasets at the 100M scale are similar to those at the 1B scale, validating the system's scalability. Compared to \textit{Shard} and \textit{Global} approaches, our system achieves 2.2- 2.8$\times$ higher throughput when recall $\ge$ 0.9. The performance gains are pronounced in all recall regimes, demonstrating CoTra's effectiveness for quality-sensitive applications.

The consistent speedup observed across all datasets and system scales further validates the generality of \sys.

\stitle{Effect of varying $k$}
To verify that \sys is capable of handling different values of $k$ in recall@k, we conduct experiments on the Text2Image dataset, which represents a challenging search scenario. Figure~\ref{fig:recall@} shows the throughput performance for $k = 1, 10, 100$, where \sys outperforms the best baseline by 2.51--2.58$\times$. Although not shown in the paper due to space limitations, similar patterns are observed across other datasets. These results confirm CoTra's consistent performance advantages under varying retrieval requirements.

\stitle{Scalability}
To demonstrate the scalability of \sys, we present performance gains in Table~\ref{tab:comparison}, where the reported QPS improvements are relative to the in-memory Single baseline with 8 threads. \sys achieves outstanding scalability, with 6.2- 7.6$\times$ improvements on 8-machines and 9.8- 13.4$\times$ QPS improvement over the single-machine system.

\sys successfully maintains near-linear scaling by leveraging the proposed techniques. Compared to the Shard index, \sys outperforms it and requires significantly fewer node accesses and computations per query to achieve equivalent recall levels. A naive application of the global graph index (\textit{Global}), even with RDMA, does not deliver ideal performance because of intensive communication. This efficiency of \sys stems from the global graph index, which effectively preserves the routing search path across the entire dataset, even in distributed settings, and enhances query traversal, as well as the design of collaborative traversal, which eliminates the redundant computations seen in partitioned approaches (e.g., Shard and Milvus) and reduces communications.

\subsection{Micro Experiments and Ablation Study}

\begin{figure}[!t]
  \centering
  \includegraphics[width=0.95\linewidth]{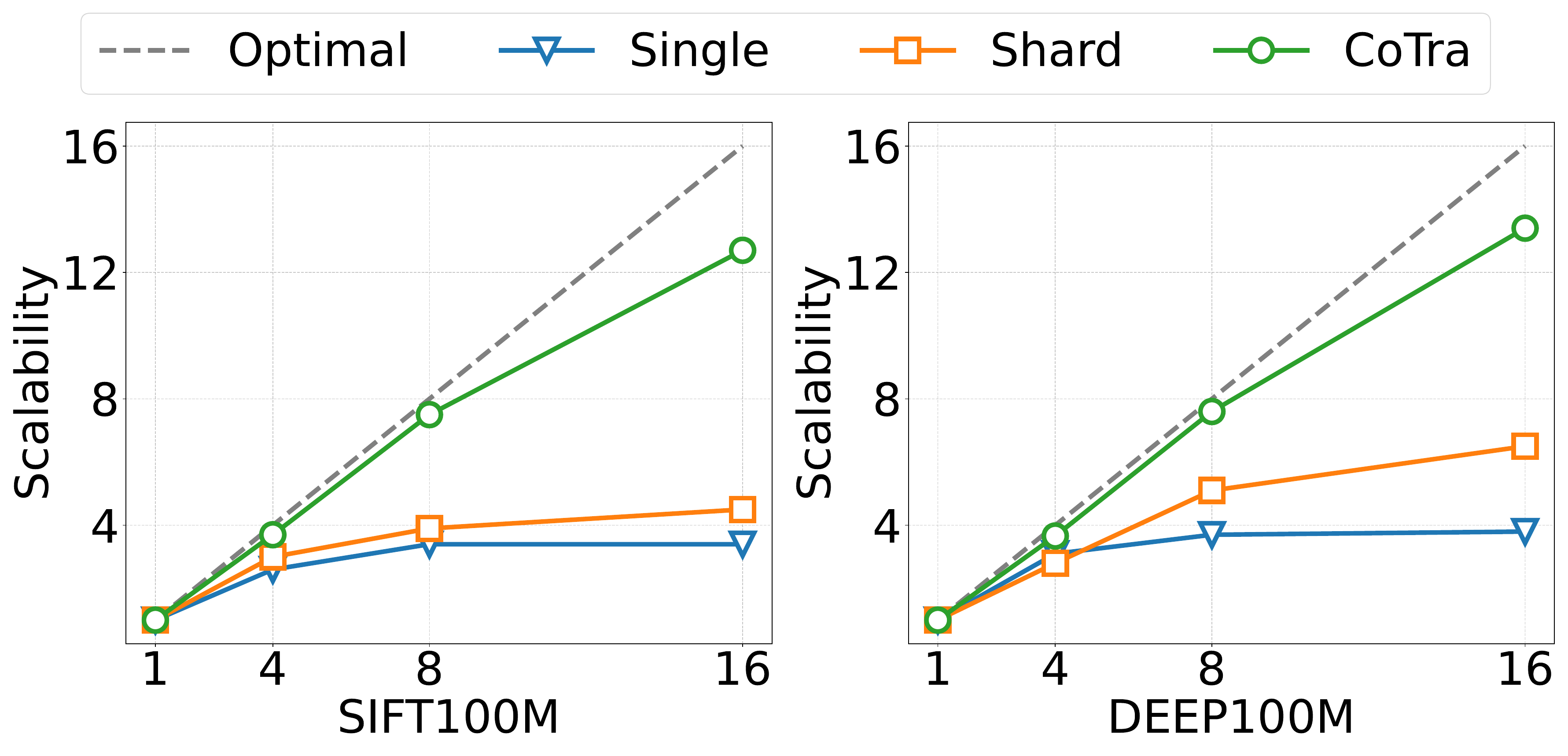}
  \caption{Scalability over machine numbers. \textit{Single} represents the total thread count scaled with the number of machines. The \textit{Global} baseline is not shown due to its scalability being below 1.}
  \label{scalability}
\end{figure}

\stitle{Search costs breakdown}
To understand Cotra's gains, we evaluated the average query computation count (computation efficiency) and the proportion of communication time in total execution time (communication efficiency) for different baselines on the SIFT100M dataset across 16 machines.

As shown in Table \ref{tab:comparison}, the single-machine system exhibits the lowest average computation count and communication overhead. However, due to memory bandwidth limitations, its throughput is severely constrained when the number of threads increases. The \textit{Global} approach achieves the second-highest computation efficiency because its search logic closely resembles that of the single-machine system. 
The \textit{Shard} suffers from increased computational operations due to partitioned index construction and search, resulting in over 4$\times$ redundant computation and thus limited throughput. In contrast, CoTra maintains relatively high computation efficiency (only $\sim$20\% redundant computation introduced by delayed updates) while achieving low communication overhead ($\sim$22\% communication time proportion), ultimately delivering the highest throughput. 

\begin{table}[!t]
\centering
\caption{Efficiency profiling of different solutions on SIFT100M. \textit{Avg. Comp.}, \textit{Comm. Ratio}, and \textit{Throughput} represent the average query computation count, communication-to-computation ratio, and QPS at 0.95 recall, respectively. }
\begin{tabular}{lccc}
\toprule
\textbf{Solution} & \textbf{Avg. Comp.} & \textbf{Comm. Ratio} & \textbf{Throughput} \\
\midrule
Single   & 3.59K  & 0.00\%    & 30.3K\\
Global   & 3.73K  & 89.4\% & 9.7K\\
Shard    & 15.6K &  <1.00\% & 40.8K\\
CoTra    & 4.33K  & 22.5\% & 116.6K\\
\bottomrule
\end{tabular}
\label{tab:comparison}
\end{table}

\stitle{Index construction time} The distributed parallel graph construction in \sys greatly accelerates index building compared to a single machine. As shown in Table~\ref{tab:index_time}, constructing the DEEP1B dataset drops from over 5 days to just 14 hours with 16 nodes. The single-machine setup uses disk-based indexing~\cite{diskann} due to out-of-memory issues during index construction  (The above search evaluations are conducted in-memory).

\begin{table}[!t]
\centering
\caption{Index construction time on each dataset.}
\scalebox{0.95}{
\begin{tabular}{cccc}
\toprule
\textbf{Dataset} & \textbf{Single} & \textbf{\sys (16 Machines)} \\
\midrule
SIFT1B &  > 96 hours & 8 hours\\
DEEP1B & > 120 hours & 14 hours\\
T2I100M& 25 hours &  4 hours\\
LAION100M & 64 hours & 9 hours\\
\bottomrule
\label{tab:index_time}
\end{tabular}
}
\end{table}

\stitle{Ablation study}
We conducted a comprehensive evaluation of the key optimizations proposed in this work, analyzing their individual contributions to throughput and latency improvements. The evaluated components include: Pull-Push mode (+PP), Co-Search mode (+CS), graph layout optimization (+GL), and query manager (+QM).

\begin{figure}[!t]
  \centering
  \includegraphics[width=1.0\linewidth]{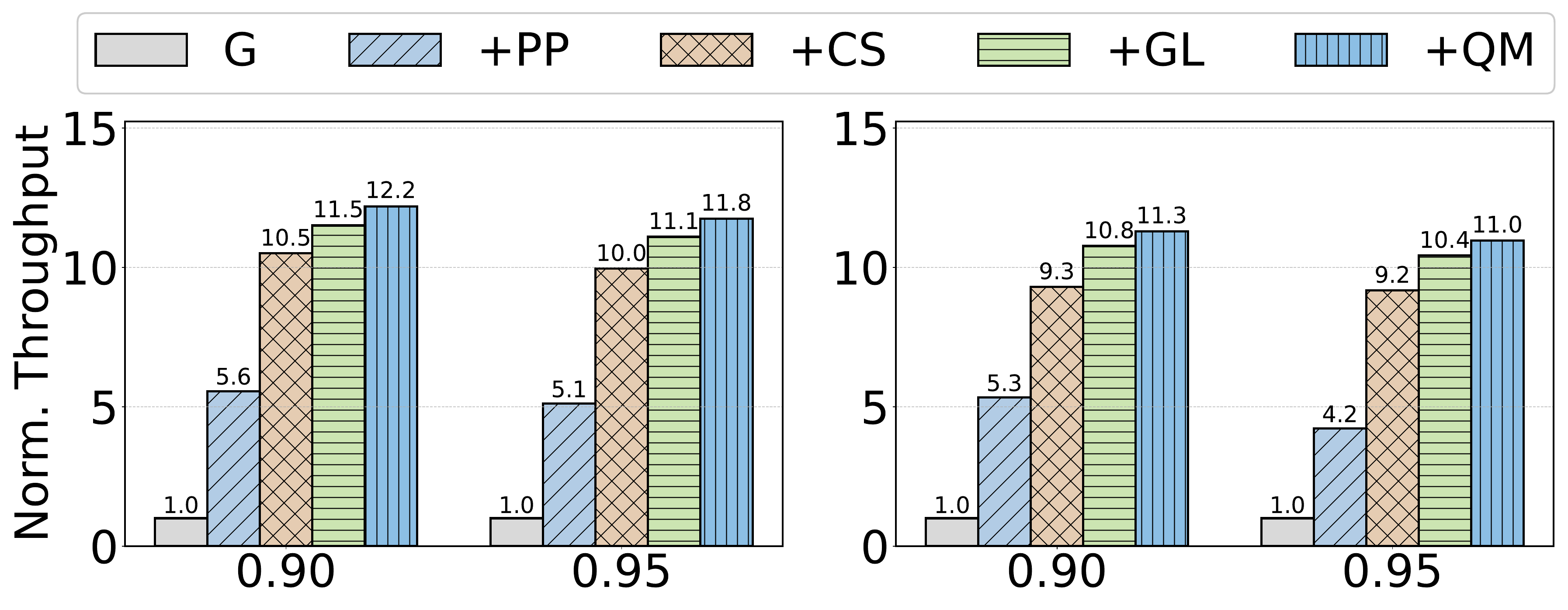}
  \caption{Speedup over \textit{Global} index at corresponding recall levels. \textbf{G} denotes the global index baseline without optimizations, \textbf{PP} represents Pull-Push mode, \textbf{CS} is Co-Search mode, \textbf{GL} indicates graph layout optimization, and \textbf{QM} stands for query manager optimization. }
  \label{fig:ablation}
\end{figure}

As shown in Figure \ref{fig:ablation}. The Pull-Push mode eliminates bulk vector transfers, the bandwidth bottleneck is significantly alleviated, leading to a substantial throughput improvement. The Co-Search mode relaxes query data dependencies, eliminating the need for passive result synchronization. This allows small batch sizes to achieve comparable QPS while drastically improving overall throughput. The optimized graph layout, combined with our Pull-Push communication mode, achieves further reduction in communication overhead. Finally, the query manager combined with other system optimizations refines computation scheduling through fine-grained workload partitioning and balancing, further improving both latency and throughput in multi-threaded scenarios.

\section{Related Work}

\stitle{In-memory systems for vector search} Vector index researches mainly consider the single-machine and in-memory setting~\cite{pq, babenko2014inverted, hnsw, nsg}. Two types of indexes are the most popular, i.e., \textit{proximity graph} that connects similar vectors to form a graph~\cite{hnsw}, and \textit{IVF} that partitions the vectors into clusters~\cite{pq}. Proximity graph requires fewer distance computations than IVF to reach the same recall~\cite{hnsw,anns_survey}, and thus our \sys utilizes the proximity graph index to enjoy its high efficiency. Many systems also optimize for in-memory query processing~\cite{sptag,parlayann,iQAN}. For instance, graph reordering is used to reduce cache miss caused by the random vector access of proximity graph in~\cite{reorder}, while cache miss is tackled with software prefetch  in~\cite{lvq}. Both ParlayANN~\cite{parlayann} and FAISS~\cite{faiss} provide efficient implementations (e.g., SIMD) for several vector indexes. 
There are also many vector quantization algorithms~\cite{rabbitQ,pq,opq,lvq} that can compress each vector into a smaller vector and accelerate distance computation. Focusing on distributed coordination, our \sys is orthogonal to these single-machine engine and vector quantization researches and can integrate them to improve performance.  

\stitle{Hybrid systems for vector search} Several systems use disk to store the large vector datasets~\cite{hmanns,smartssd}. For instance, DiskANN~\cite{diskann} keeps the compressed vectors in memory and the exact vectors and proximity graph index on disk. Starling~\cite{starling} reduces the disk read amplification of DiskANN with graph reordering. SPANN uses the IVF index by keeping the cluster centers in memory and the vectors on disk~\cite{spann}. These disk-based systems are cost-effective, but throughput is limited by disk bandwidth. Some systems use CPU-GPU joint processing to benefit from the strong computational power of GPUs~\cite {ootomo2024cagra}. For instance, Rummy uses the IVF index and moves clusters from CPU to GPU on demand for computation~\cite{rummy}, while Bang uses the proximity graph index and moves the adjacency lists from CPU to GPU to compute approximate distances~\cite{khan2024bang}. For these systems, small GPU memory and  PCIe bandwidth are major limiting factors.

\stitle{Distributed systems for vector search} Most systems conduct distributed vector search by building an independent index on each machine, e.g., Milvus~\cite{milvus}, AnalyticDB~\cite{AnalyticDB}, Weaviate~\cite{Dilocker_Weaviate},  Cassandra~\cite{Cassandra}, and Pyramid~\cite{pyramid}. This simplifies system design but degrades computation efficiency, as shown by our profiling. 
Some systems conduct distributed vector using the IVF index, e.g., Auncel~\cite{Auncel} and SPFresh ~\cite{spfresh}. The benefit is that each machine only needs to scan its local clusters after the top-ranking clusters have been decided. However, the computation efficiency of the IVF index is lower than that of the proximity graph. Some systems use new hardware to scale up vector search, e.g., CXL-ANNS~\cite{cxlanns} adopts the Compute Express Link (CXL) as a memory pool, and DF-GAS \cite{dfgasdistrifpga}  utilizes FPGAs for distance computation. To our knowledge, \sys is the first to conduct distributed vector using a holistic proximity graph index over the entire dataset. This enjoys high computation efficiency but requires careful algorithm-system co-designs to handle the vast communications induced by the fine-grained data dependencies of the proximity graph index.


\section{Conclusions}

We present \sys, a system that scales up vector search over multiple machines. We find that using a holistic proximity graph index for the entire dataset enjoys high computation efficiency but suffers from the extensive communication caused by rich data dependencies. To reduce the communication cost, we conduct algorithm system co-designs by observing that the efficiency of proximity graph search can be maintained if the quality of the candidates are good. We also incorporate system optimizations including task scheduling, RDMA-friendly graph format optimization, and distributed index building. To our knowledge, \sys is the first system that achieves close to linear throughput scaling compared to proximity graph on a single machine.

\bibliographystyle{ACM-Reference-Format}
\bibliography{acmart}


\begin{thebibliography}{65}


\ifx \showCODEN    \undefined \def \showCODEN     #1{\unskip}     \fi
\ifx \showISBNx    \undefined \def \showISBNx     #1{\unskip}     \fi
\ifx \showISBNxiii \undefined \def \showISBNxiii  #1{\unskip}     \fi
\ifx \showISSN     \undefined \def \showISSN      #1{\unskip}     \fi
\ifx \showLCCN     \undefined \def \showLCCN      #1{\unskip}     \fi
\ifx \shownote     \undefined \def \shownote      #1{#1}          \fi
\ifx \showarticletitle \undefined \def \showarticletitle #1{#1}   \fi
\ifx \showURL      \undefined \def \showURL       {\relax}        \fi
\providecommand\bibfield[2]{#2}
\providecommand\bibinfo[2]{#2}
\providecommand\natexlab[1]{#1}
\providecommand\showeprint[2][]{arXiv:#2}

\bibitem[Aguerrebere et~al\mbox{.}(2023)]%
        {lvq}
\bibfield{author}{\bibinfo{person}{Cecilia Aguerrebere}, \bibinfo{person}{Ishwar~Singh Bhati}, \bibinfo{person}{Mark Hildebrand}, \bibinfo{person}{Mariano Tepper}, {and} \bibinfo{person}{Theodore Willke}.} \bibinfo{year}{2023}\natexlab{}.
\newblock \showarticletitle{Similarity Search in the Blink of an Eye with Compressed Indices}.
\newblock \bibinfo{journal}{\emph{Proc. VLDB Endow.}} \bibinfo{volume}{16}, \bibinfo{number}{11} (\bibinfo{date}{July} \bibinfo{year}{2023}), \bibinfo{pages}{3433–3446}.
\newblock
\showISSN{2150-8097}
\href{https://doi.org/10.14778/3611479.3611537}{doi:\nolinkurl{10.14778/3611479.3611537}}


\bibitem[Babenko and Lempitsky(2014)]%
        {babenko2014inverted}
\bibfield{author}{\bibinfo{person}{Artem Babenko} {and} \bibinfo{person}{Victor Lempitsky}.} \bibinfo{year}{2014}\natexlab{}.
\newblock \showarticletitle{The inverted multi-index}.
\newblock \bibinfo{journal}{\emph{IEEE transactions on pattern analysis and machine intelligence}} \bibinfo{volume}{37}, \bibinfo{number}{6} (\bibinfo{year}{2014}), \bibinfo{pages}{1247--1260}.
\newblock


\bibitem[Babenko and Lempitsky(2016)]%
        {babenko2016efficient}
\bibfield{author}{\bibinfo{person}{Artem Babenko} {and} \bibinfo{person}{Victor Lempitsky}.} \bibinfo{year}{2016}\natexlab{}.
\newblock \showarticletitle{Efficient indexing of billion-scale datasets of deep descriptors}. In \bibinfo{booktitle}{\emph{Proceedings of the IEEE Conference on Computer Vision and Pattern Recognition}}. \bibinfo{pages}{2055--2063}.
\newblock


\bibitem[Bittremieux et~al\mbox{.}(2018)]%
        {bittremieux2018fast}
\bibfield{author}{\bibinfo{person}{Wout Bittremieux}, \bibinfo{person}{Pieter Meysman}, \bibinfo{person}{William~Stafford Noble}, {and} \bibinfo{person}{Kris Laukens}.} \bibinfo{year}{2018}\natexlab{}.
\newblock \showarticletitle{Fast open modification spectral library searching through approximate nearest neighbor indexing}.
\newblock \bibinfo{journal}{\emph{Journal of proteome research}} \bibinfo{volume}{17}, \bibinfo{number}{10} (\bibinfo{year}{2018}), \bibinfo{pages}{3463--3474}.
\newblock


\bibitem[Chen et~al\mbox{.}(2024)]%
        {chen2024bge}
\bibfield{author}{\bibinfo{person}{Jianlv Chen}, \bibinfo{person}{Shitao Xiao}, \bibinfo{person}{Peitian Zhang}, \bibinfo{person}{Kun Luo}, \bibinfo{person}{Defu Lian}, {and} \bibinfo{person}{Zheng Liu}.} \bibinfo{year}{2024}\natexlab{}.
\newblock \showarticletitle{Bge m3-embedding: Multi-lingual, multi-functionality, multi-granularity text embeddings through self-knowledge distillation}.
\newblock \bibinfo{journal}{\emph{arXiv preprint arXiv:2402.03216}} (\bibinfo{year}{2024}).
\newblock


\bibitem[Chen et~al\mbox{.}(2018)]%
        {sptag}
\bibfield{author}{\bibinfo{person}{Qi Chen}, \bibinfo{person}{Haidong Wang}, \bibinfo{person}{Mingqin Li}, \bibinfo{person}{Gang Ren}, \bibinfo{person}{Scarlett Li}, \bibinfo{person}{Jeffery Zhu}, \bibinfo{person}{Jason Li}, \bibinfo{person}{Chuanjie Liu}, \bibinfo{person}{Lintao Zhang}, {and} \bibinfo{person}{Jingdong Wang}.} \bibinfo{year}{2018}\natexlab{}.
\newblock \bibinfo{booktitle}{\emph{SPTAG: A library for fast approximate nearest neighbor search}}.
\newblock
\urldef\tempurl%
\url{https://github.com/Microsoft/SPTAG}
\showURL{%
\tempurl}


\bibitem[Chen et~al\mbox{.}(2021)]%
        {spann}
\bibfield{author}{\bibinfo{person}{Qi Chen}, \bibinfo{person}{Bing Zhao}, \bibinfo{person}{Haidong Wang}, \bibinfo{person}{Mingqin Li}, \bibinfo{person}{Chuanjie Liu}, \bibinfo{person}{Zengzhong Li}, \bibinfo{person}{Mao Yang}, {and} \bibinfo{person}{Jingdong Wang}.} \bibinfo{year}{2021}\natexlab{}.
\newblock \showarticletitle{SPANN: Highly-efficient Billion-scale Approximate Nearest Neighborhood Search}. In \bibinfo{booktitle}{\emph{Advances in Neural Information Processing Systems}}, \bibfield{editor}{\bibinfo{person}{M.~Ranzato}, \bibinfo{person}{A.~Beygelzimer}, \bibinfo{person}{Y.~Dauphin}, \bibinfo{person}{P.S. Liang}, {and} \bibinfo{person}{J.~Wortman Vaughan}} (Eds.), Vol.~\bibinfo{volume}{34}. \bibinfo{publisher}{Curran Associates, Inc.}, \bibinfo{pages}{5199--5212}.
\newblock
\urldef\tempurl%
\url{https://proceedings.neurips.cc/paper_files/paper/2021/file/299dc35e747eb77177d9cea10a802da2-Paper.pdf}
\showURL{%
\tempurl}


\bibitem[Chen and Zou(2024)]%
        {chen2024genept}
\bibfield{author}{\bibinfo{person}{Yiqun Chen} {and} \bibinfo{person}{James Zou}.} \bibinfo{year}{2024}\natexlab{}.
\newblock \showarticletitle{GenePT: a simple but effective foundation model for genes and cells built from ChatGPT}.
\newblock \bibinfo{journal}{\emph{bioRxiv}} (\bibinfo{year}{2024}), \bibinfo{pages}{2023--10}.
\newblock


\bibitem[Coleman et~al\mbox{.}(2022)]%
        {reorder}
\bibfield{author}{\bibinfo{person}{Benjamin Coleman}, \bibinfo{person}{Santiago Segarra}, \bibinfo{person}{Alexander~J Smola}, {and} \bibinfo{person}{Anshumali Shrivastava}.} \bibinfo{year}{2022}\natexlab{}.
\newblock \showarticletitle{Graph Reordering for Cache-Efficient Near Neighbor Search}. In \bibinfo{booktitle}{\emph{Advances in Neural Information Processing Systems}}, \bibfield{editor}{\bibinfo{person}{S.~Koyejo}, \bibinfo{person}{S.~Mohamed}, \bibinfo{person}{A.~Agarwal}, \bibinfo{person}{D.~Belgrave}, \bibinfo{person}{K.~Cho}, {and} \bibinfo{person}{A.~Oh}} (Eds.), Vol.~\bibinfo{volume}{35}. \bibinfo{publisher}{Curran Associates, Inc.}, \bibinfo{pages}{38488--38500}.
\newblock
\urldef\tempurl%
\url{https://proceedings.neurips.cc/paper_files/paper/2022/file/fb44a668c2d4bc984e9d6ca261262cbb-Paper-Conference.pdf}
\showURL{%
\tempurl}


\bibitem[Covington et~al\mbox{.}(2016)]%
        {covington2016deep}
\bibfield{author}{\bibinfo{person}{Paul Covington}, \bibinfo{person}{Jay Adams}, {and} \bibinfo{person}{Emre Sargin}.} \bibinfo{year}{2016}\natexlab{}.
\newblock \showarticletitle{Deep neural networks for youtube recommendations}. In \bibinfo{booktitle}{\emph{Proceedings of the 10th ACM conference on recommender systems}}. \bibinfo{pages}{191--198}.
\newblock


\bibitem[Deng et~al\mbox{.}(2019)]%
        {pyramid}
\bibfield{author}{\bibinfo{person}{Shiyuan Deng}, \bibinfo{person}{Xiao Yan}, \bibinfo{person}{K.W.~Ng Kelvin}, \bibinfo{person}{Chenyu Jiang}, {and} \bibinfo{person}{James Cheng}.} \bibinfo{year}{2019}\natexlab{}.
\newblock \showarticletitle{Pyramid: A General Framework for Distributed Similarity Search on Large-scale Datasets}. In \bibinfo{booktitle}{\emph{2019 IEEE International Conference on Big Data (Big Data)}}. \bibinfo{pages}{1066--1071}.
\newblock
\href{https://doi.org/10.1109/BigData47090.2019.9006219}{doi:\nolinkurl{10.1109/BigData47090.2019.9006219}}


\bibitem[Dilocker et~al\mbox{.}({[n.\,d.]})]%
        {Dilocker_Weaviate}
\bibfield{author}{\bibinfo{person}{Etienne Dilocker}, \bibinfo{person}{Bob van Luijt}, \bibinfo{person}{Byron Voorbach}, \bibinfo{person}{Mohd~Shukri Hasan}, \bibinfo{person}{Abdel Rodriguez}, \bibinfo{person}{Dirk~Alexander Kulawiak}, \bibinfo{person}{Marcin Antas}, {and} \bibinfo{person}{Parker Duckworth}.} \bibinfo{year}{[n.\,d.]}\natexlab{}.
\newblock \bibinfo{booktitle}{\emph{{Weaviate}}}.
\newblock
\urldef\tempurl%
\url{https://github.com/weaviate/weaviate}
\showURL{%
\tempurl}


\bibitem[Dmitry~Baranchuk(2021)]%
        {yandextexttoimage}
\bibfield{author}{\bibinfo{person}{Artem~Babenko Dmitry~Baranchuk}.} \bibinfo{year}{2021}\natexlab{}.
\newblock \bibinfo{title}{Text-to-Image dataset for billion-scale similarity search}.
\newblock
\urldef\tempurl%
\url{https://research.yandex.com/datasets/text-to-image-dataset-for-billion-scale-similarity-search}
\showURL{%
Retrieved April 13, 2025 from \tempurl}


\bibitem[Fent et~al\mbox{.}(2020)]%
        {fent2020low}
\bibfield{author}{\bibinfo{person}{Philipp Fent}, \bibinfo{person}{Alexander van Renen}, \bibinfo{person}{Andreas Kipf}, \bibinfo{person}{Viktor Leis}, \bibinfo{person}{Thomas Neumann}, {and} \bibinfo{person}{Alfons Kemper}.} \bibinfo{year}{2020}\natexlab{}.
\newblock \showarticletitle{Low-latency communication for fast DBMS using RDMA and shared memory}. In \bibinfo{booktitle}{\emph{2020 IEEE 36th International Conference on Data Engineering (ICDE)}}. IEEE, \bibinfo{pages}{1477--1488}.
\newblock


\bibitem[Fu et~al\mbox{.}(2019)]%
        {nsg}
\bibfield{author}{\bibinfo{person}{Cong Fu}, \bibinfo{person}{Chao Xiang}, \bibinfo{person}{Changxu Wang}, {and} \bibinfo{person}{Deng Cai}.} \bibinfo{year}{2019}\natexlab{}.
\newblock \showarticletitle{Fast approximate nearest neighbor search with the navigating spreading-out graph}.
\newblock \bibinfo{journal}{\emph{Proc. VLDB Endow.}} \bibinfo{volume}{12}, \bibinfo{number}{5} (\bibinfo{date}{Jan.} \bibinfo{year}{2019}), \bibinfo{pages}{461–474}.
\newblock
\showISSN{2150-8097}
\href{https://doi.org/10.14778/3303753.3303754}{doi:\nolinkurl{10.14778/3303753.3303754}}


\bibitem[Gao and Long(2024)]%
        {rabbitQ}
\bibfield{author}{\bibinfo{person}{Jianyang Gao} {and} \bibinfo{person}{Cheng Long}.} \bibinfo{year}{2024}\natexlab{}.
\newblock \showarticletitle{RaBitQ: Quantizing High-Dimensional Vectors with a Theoretical Error Bound for Approximate Nearest Neighbor Search}.
\newblock \bibinfo{journal}{\emph{Proc. ACM Manag. Data}} \bibinfo{volume}{2}, \bibinfo{number}{3}, Article \bibinfo{articleno}{167} (\bibinfo{date}{May} \bibinfo{year}{2024}), \bibinfo{numpages}{27}~pages.
\newblock
\href{https://doi.org/10.1145/3654970}{doi:\nolinkurl{10.1145/3654970}}


\bibitem[Ge et~al\mbox{.}(2014)]%
        {opq}
\bibfield{author}{\bibinfo{person}{Tiezheng Ge}, \bibinfo{person}{Kaiming He}, \bibinfo{person}{Qifa Ke}, {and} \bibinfo{person}{Jian Sun}.} \bibinfo{year}{2014}\natexlab{}.
\newblock \showarticletitle{Optimized Product Quantization}.
\newblock \bibinfo{journal}{\emph{IEEE Transactions on Pattern Analysis and Machine Intelligence}} \bibinfo{volume}{36}, \bibinfo{number}{4} (\bibinfo{year}{2014}), \bibinfo{pages}{744--755}.
\newblock
\href{https://doi.org/10.1109/TPAMI.2013.240}{doi:\nolinkurl{10.1109/TPAMI.2013.240}}


\bibitem[Huang et~al\mbox{.}(2020)]%
        {huang2020embedding}
\bibfield{author}{\bibinfo{person}{Jui-Ting Huang}, \bibinfo{person}{Ashish Sharma}, \bibinfo{person}{Shuying Sun}, \bibinfo{person}{Li Xia}, \bibinfo{person}{David Zhang}, \bibinfo{person}{Philip Pronin}, \bibinfo{person}{Janani Padmanabhan}, \bibinfo{person}{Giuseppe Ottaviano}, {and} \bibinfo{person}{Linjun Yang}.} \bibinfo{year}{2020}\natexlab{}.
\newblock \showarticletitle{Embedding-based retrieval in facebook search}. In \bibinfo{booktitle}{\emph{Proceedings of the 26th ACM SIGKDD International Conference on Knowledge Discovery \& Data Mining}}. \bibinfo{pages}{2553--2561}.
\newblock


\bibitem[Indyk and Motwani(1998)]%
        {approximate_graph}
\bibfield{author}{\bibinfo{person}{Piotr Indyk} {and} \bibinfo{person}{Rajeev Motwani}.} \bibinfo{year}{1998}\natexlab{}.
\newblock \showarticletitle{Approximate nearest neighbors: towards removing the curse of dimensionality}. In \bibinfo{booktitle}{\emph{Proceedings of the Thirtieth Annual ACM Symposium on Theory of Computing}} (Dallas, Texas, USA) \emph{(\bibinfo{series}{STOC '98})}. \bibinfo{publisher}{Association for Computing Machinery}, \bibinfo{address}{New York, NY, USA}, \bibinfo{pages}{604–613}.
\newblock
\showISBNx{0897919629}
\href{https://doi.org/10.1145/276698.276876}{doi:\nolinkurl{10.1145/276698.276876}}


\bibitem[Jang et~al\mbox{.}(2023)]%
        {cxlanns}
\bibfield{author}{\bibinfo{person}{Junhyeok Jang}, \bibinfo{person}{Hanjin Choi}, \bibinfo{person}{Hanyeoreum Bae}, \bibinfo{person}{Seungjun Lee}, \bibinfo{person}{Miryeong Kwon}, {and} \bibinfo{person}{Myoungsoo Jung}.} \bibinfo{year}{2023}\natexlab{}.
\newblock \showarticletitle{{CXL-ANNS}: {Software-Hardware} Collaborative Memory Disaggregation and Computation for {Billion-Scale} Approximate Nearest Neighbor Search}. In \bibinfo{booktitle}{\emph{2023 USENIX Annual Technical Conference (USENIX ATC 23)}}. \bibinfo{publisher}{USENIX Association}, \bibinfo{address}{Boston, MA}, \bibinfo{pages}{585--600}.
\newblock
\showISBNx{978-1-939133-35-9}
\urldef\tempurl%
\url{https://www.usenix.org/conference/atc23/presentation/jang}
\showURL{%
\tempurl}


\bibitem[Jegou et~al\mbox{.}(2010)]%
        {jegou2010product}
\bibfield{author}{\bibinfo{person}{Herve Jegou}, \bibinfo{person}{Matthijs Douze}, {and} \bibinfo{person}{Cordelia Schmid}.} \bibinfo{year}{2010}\natexlab{}.
\newblock \showarticletitle{Product quantization for nearest neighbor search}.
\newblock \bibinfo{journal}{\emph{IEEE transactions on pattern analysis and machine intelligence}} \bibinfo{volume}{33}, \bibinfo{number}{1} (\bibinfo{year}{2010}), \bibinfo{pages}{117--128}.
\newblock


\bibitem[Jiang et~al\mbox{.}(2024)]%
        {metascale}
\bibfield{author}{\bibinfo{person}{Ziheng Jiang}, \bibinfo{person}{Haibin Lin}, \bibinfo{person}{Yinmin Zhong}, \bibinfo{person}{Qi Huang}, \bibinfo{person}{Yangrui Chen}, \bibinfo{person}{Zhi Zhang}, \bibinfo{person}{Yanghua Peng}, \bibinfo{person}{Xiang Li}, \bibinfo{person}{Cong Xie}, \bibinfo{person}{Shibiao Nong}, \bibinfo{person}{Yulu Jia}, \bibinfo{person}{Sun He}, \bibinfo{person}{Hongmin Chen}, \bibinfo{person}{Zhihao Bai}, \bibinfo{person}{Qi Hou}, \bibinfo{person}{Shipeng Yan}, \bibinfo{person}{Ding Zhou}, \bibinfo{person}{Yiyao Sheng}, \bibinfo{person}{Zhuo Jiang}, \bibinfo{person}{Haohan Xu}, \bibinfo{person}{Haoran Wei}, \bibinfo{person}{Zhang Zhang}, \bibinfo{person}{Pengfei Nie}, \bibinfo{person}{Leqi Zou}, \bibinfo{person}{Sida Zhao}, \bibinfo{person}{Liang Xiang}, \bibinfo{person}{Zherui Liu}, \bibinfo{person}{Zhe Li}, \bibinfo{person}{Xiaoying Jia}, \bibinfo{person}{Jianxi Ye}, \bibinfo{person}{Xin Jin}, {and} \bibinfo{person}{Xin Liu}.} \bibinfo{year}{2024}\natexlab{}.
\newblock \showarticletitle{{MegaScale}: Scaling Large Language Model Training to More Than 10,000 {GPUs}}. In \bibinfo{booktitle}{\emph{21st USENIX Symposium on Networked Systems Design and Implementation (NSDI 24)}}. \bibinfo{publisher}{USENIX Association}, \bibinfo{address}{Santa Clara, CA}, \bibinfo{pages}{745--760}.
\newblock
\showISBNx{978-1-939133-39-7}
\urldef\tempurl%
\url{https://www.usenix.org/conference/nsdi24/presentation/jiang-ziheng}
\showURL{%
\tempurl}


\bibitem[Johnson et~al\mbox{.}(2021)]%
        {faiss}
\bibfield{author}{\bibinfo{person}{Jeff Johnson}, \bibinfo{person}{Matthijs Douze}, {and} \bibinfo{person}{Hervé Jégou}.} \bibinfo{year}{2021}\natexlab{}.
\newblock \showarticletitle{Billion-Scale Similarity Search with GPUs}.
\newblock \bibinfo{journal}{\emph{IEEE Transactions on Big Data}} \bibinfo{volume}{7}, \bibinfo{number}{3} (\bibinfo{year}{2021}), \bibinfo{pages}{535--547}.
\newblock
\href{https://doi.org/10.1109/TBDATA.2019.2921572}{doi:\nolinkurl{10.1109/TBDATA.2019.2921572}}


\bibitem[Jégou et~al\mbox{.}(2011)]%
        {pq}
\bibfield{author}{\bibinfo{person}{Herve Jégou}, \bibinfo{person}{Matthijs Douze}, {and} \bibinfo{person}{Cordelia Schmid}.} \bibinfo{year}{2011}\natexlab{}.
\newblock \showarticletitle{Product Quantization for Nearest Neighbor Search}.
\newblock \bibinfo{journal}{\emph{IEEE Transactions on Pattern Analysis and Machine Intelligence}} \bibinfo{volume}{33}, \bibinfo{number}{1} (\bibinfo{year}{2011}), \bibinfo{pages}{117--128}.
\newblock
\href{https://doi.org/10.1109/TPAMI.2010.57}{doi:\nolinkurl{10.1109/TPAMI.2010.57}}


\bibitem[Kalantidis and Avrithis(2014)]%
        {kalantidis2014locally}
\bibfield{author}{\bibinfo{person}{Yannis Kalantidis} {and} \bibinfo{person}{Yannis Avrithis}.} \bibinfo{year}{2014}\natexlab{}.
\newblock \showarticletitle{Locally optimized product quantization for approximate nearest neighbor search}. In \bibinfo{booktitle}{\emph{Proceedings of the IEEE conference on computer vision and pattern recognition}}. \bibinfo{pages}{2321--2328}.
\newblock


\bibitem[Kalia et~al\mbox{.}(2014)]%
        {kalia2014using}
\bibfield{author}{\bibinfo{person}{Anuj Kalia}, \bibinfo{person}{Michael Kaminsky}, {and} \bibinfo{person}{David~G Andersen}.} \bibinfo{year}{2014}\natexlab{}.
\newblock \showarticletitle{Using RDMA efficiently for key-value services}. In \bibinfo{booktitle}{\emph{Proceedings of the 2014 ACM Conference on SIGCOMM}}. \bibinfo{pages}{295--306}.
\newblock


\bibitem[Kalia et~al\mbox{.}(2016)]%
        {fasst}
\bibfield{author}{\bibinfo{person}{Anuj Kalia}, \bibinfo{person}{Michael Kaminsky}, {and} \bibinfo{person}{David~G. Andersen}.} \bibinfo{year}{2016}\natexlab{}.
\newblock \showarticletitle{{FaSST}: Fast, Scalable and Simple Distributed Transactions with {Two-Sided} ({{{{{RDMA}}}}}) Datagram {RPCs}}. In \bibinfo{booktitle}{\emph{12th USENIX Symposium on Operating Systems Design and Implementation (OSDI 16)}}. \bibinfo{publisher}{USENIX Association}, \bibinfo{address}{Savannah, GA}, \bibinfo{pages}{185--201}.
\newblock
\showISBNx{978-1-931971-33-1}
\urldef\tempurl%
\url{https://www.usenix.org/conference/osdi16/technical-sessions/presentation/kalia}
\showURL{%
\tempurl}


\bibitem[Khan et~al\mbox{.}(2024)]%
        {khan2024bang}
\bibfield{author}{\bibinfo{person}{Saim Khan}, \bibinfo{person}{Somesh Singh}, \bibinfo{person}{Harsha~Vardhan Simhadri}, \bibinfo{person}{Jyothi Vedurada}, {et~al\mbox{.}}} \bibinfo{year}{2024}\natexlab{}.
\newblock \showarticletitle{BANG: Billion-Scale Approximate Nearest Neighbor Search using a Single GPU}.
\newblock \bibinfo{journal}{\emph{arXiv preprint arXiv:2401.11324}} (\bibinfo{year}{2024}).
\newblock


\bibitem[Kiselev et~al\mbox{.}(2018)]%
        {kiselev2018scmap}
\bibfield{author}{\bibinfo{person}{Vladimir~Yu Kiselev}, \bibinfo{person}{Andrew Yiu}, {and} \bibinfo{person}{Martin Hemberg}.} \bibinfo{year}{2018}\natexlab{}.
\newblock \showarticletitle{scmap: projection of single-cell RNA-seq data across data sets}.
\newblock \bibinfo{journal}{\emph{Nature methods}} \bibinfo{volume}{15}, \bibinfo{number}{5} (\bibinfo{year}{2018}), \bibinfo{pages}{359--362}.
\newblock


\bibitem[Lakshman and Malik(2010)]%
        {Cassandra}
\bibfield{author}{\bibinfo{person}{Avinash Lakshman} {and} \bibinfo{person}{Prashant Malik}.} \bibinfo{year}{2010}\natexlab{}.
\newblock \showarticletitle{Cassandra: a decentralized structured storage system}.
\newblock \bibinfo{journal}{\emph{SIGOPS Oper. Syst. Rev.}} \bibinfo{volume}{44}, \bibinfo{number}{2} (\bibinfo{date}{April} \bibinfo{year}{2010}), \bibinfo{pages}{35–40}.
\newblock
\showISSN{0163-5980}
\href{https://doi.org/10.1145/1773912.1773922}{doi:\nolinkurl{10.1145/1773912.1773922}}


\bibitem[Laurent~Amsaleg(2010)]%
        {sift}
\bibfield{author}{\bibinfo{person}{Hervé~Jégou Laurent~Amsaleg}.} \bibinfo{year}{2010}\natexlab{}.
\newblock \bibinfo{title}{Datasets for approximate nearest neighbor search}.
\newblock
\urldef\tempurl%
\url{http://corpus-texmex.irisa.fr/}
\showURL{%
\tempurl}
\newblock
\shownote{Accessed: 2025-04-17}.


\bibitem[Li et~al\mbox{.}(2019)]%
        {li2019approximate}
\bibfield{author}{\bibinfo{person}{Wen Li}, \bibinfo{person}{Ying Zhang}, \bibinfo{person}{Yifang Sun}, \bibinfo{person}{Wei Wang}, \bibinfo{person}{Mingjie Li}, \bibinfo{person}{Wenjie Zhang}, {and} \bibinfo{person}{Xuemin Lin}.} \bibinfo{year}{2019}\natexlab{}.
\newblock \showarticletitle{Approximate nearest neighbor search on high dimensional data—experiments, analyses, and improvement}.
\newblock \bibinfo{journal}{\emph{IEEE Transactions on Knowledge and Data Engineering}} \bibinfo{volume}{32}, \bibinfo{number}{8} (\bibinfo{year}{2019}), \bibinfo{pages}{1475--1488}.
\newblock


\bibitem[Malkov and Yashunin(2020)]%
        {hnsw}
\bibfield{author}{\bibinfo{person}{Yu~A. Malkov} {and} \bibinfo{person}{D.~A. Yashunin}.} \bibinfo{year}{2020}\natexlab{}.
\newblock \showarticletitle{Efficient and Robust Approximate Nearest Neighbor Search Using Hierarchical Navigable Small World Graphs}.
\newblock \bibinfo{journal}{\emph{IEEE Transactions on Pattern Analysis and Machine Intelligence}} \bibinfo{volume}{42}, \bibinfo{number}{4} (\bibinfo{year}{2020}), \bibinfo{pages}{824--836}.
\newblock
\href{https://doi.org/10.1109/TPAMI.2018.2889473}{doi:\nolinkurl{10.1109/TPAMI.2018.2889473}}


\bibitem[Manohar et~al\mbox{.}(2024)]%
        {parlayann}
\bibfield{author}{\bibinfo{person}{Magdalen~Dobson Manohar}, \bibinfo{person}{Zheqi Shen}, \bibinfo{person}{Guy Blelloch}, \bibinfo{person}{Laxman Dhulipala}, \bibinfo{person}{Yan Gu}, \bibinfo{person}{Harsha~Vardhan Simhadri}, {and} \bibinfo{person}{Yihan Sun}.} \bibinfo{year}{2024}\natexlab{}.
\newblock \showarticletitle{ParlayANN: Scalable and Deterministic Parallel Graph-Based Approximate Nearest Neighbor Search Algorithms}. In \bibinfo{booktitle}{\emph{Proceedings of the 29th ACM SIGPLAN Annual Symposium on Principles and Practice of Parallel Programming}} (Edinburgh, United Kingdom) \emph{(\bibinfo{series}{PPoPP '24})}. \bibinfo{publisher}{Association for Computing Machinery}, \bibinfo{address}{New York, NY, USA}, \bibinfo{pages}{270–285}.
\newblock
\showISBNx{9798400704352}
\href{https://doi.org/10.1145/3627535.3638475}{doi:\nolinkurl{10.1145/3627535.3638475}}


\bibitem[Nigam et~al\mbox{.}(2019)]%
        {nigam2019semantic}
\bibfield{author}{\bibinfo{person}{Priyanka Nigam}, \bibinfo{person}{Yiwei Song}, \bibinfo{person}{Vijai Mohan}, \bibinfo{person}{Vihan Lakshman}, \bibinfo{person}{Weitian Ding}, \bibinfo{person}{Ankit Shingavi}, \bibinfo{person}{Choon~Hui Teo}, \bibinfo{person}{Hao Gu}, {and} \bibinfo{person}{Bing Yin}.} \bibinfo{year}{2019}\natexlab{}.
\newblock \showarticletitle{Semantic product search}. In \bibinfo{booktitle}{\emph{Proceedings of the 25th ACM SIGKDD International Conference on Knowledge Discovery \& Data Mining}}. \bibinfo{pages}{2876--2885}.
\newblock


\bibitem[Ootomo et~al\mbox{.}(2024)]%
        {ootomo2024cagra}
\bibfield{author}{\bibinfo{person}{Hiroyuki Ootomo}, \bibinfo{person}{Akira Naruse}, \bibinfo{person}{Corey Nolet}, \bibinfo{person}{Ray Wang}, \bibinfo{person}{Tamas Feher}, {and} \bibinfo{person}{Yong Wang}.} \bibinfo{year}{2024}\natexlab{}.
\newblock \showarticletitle{Cagra: Highly parallel graph construction and approximate nearest neighbor search for gpus}. In \bibinfo{booktitle}{\emph{2024 IEEE 40th International Conference on Data Engineering (ICDE)}}. IEEE, \bibinfo{pages}{4236--4247}.
\newblock


\bibitem[Peng et~al\mbox{.}(2023a)]%
        {peng2023efficient}
\bibfield{author}{\bibinfo{person}{Yun Peng}, \bibinfo{person}{Byron Choi}, \bibinfo{person}{Tsz~Nam Chan}, \bibinfo{person}{Jianye Yang}, {and} \bibinfo{person}{Jianliang Xu}.} \bibinfo{year}{2023}\natexlab{a}.
\newblock \showarticletitle{Efficient approximate nearest neighbor search in multi-dimensional databases}.
\newblock \bibinfo{journal}{\emph{Proceedings of the ACM on Management of Data}} \bibinfo{volume}{1}, \bibinfo{number}{1} (\bibinfo{year}{2023}), \bibinfo{pages}{1--27}.
\newblock


\bibitem[Peng et~al\mbox{.}(2023b)]%
        {iQAN}
\bibfield{author}{\bibinfo{person}{Zhen Peng}, \bibinfo{person}{Minjia Zhang}, \bibinfo{person}{Kai Li}, \bibinfo{person}{Ruoming Jin}, {and} \bibinfo{person}{Bin Ren}.} \bibinfo{year}{2023}\natexlab{b}.
\newblock \showarticletitle{iQAN: Fast and Accurate Vector Search with Efficient Intra-Query Parallelism on Multi-Core Architectures}. In \bibinfo{booktitle}{\emph{Proceedings of the 28th ACM SIGPLAN Annual Symposium on Principles and Practice of Parallel Programming}} (Montreal, QC, Canada) \emph{(\bibinfo{series}{PPoPP '23})}. \bibinfo{publisher}{Association for Computing Machinery}, \bibinfo{address}{New York, NY, USA}, \bibinfo{pages}{313–328}.
\newblock
\showISBNx{9798400700156}
\href{https://doi.org/10.1145/3572848.3577527}{doi:\nolinkurl{10.1145/3572848.3577527}}


\bibitem[Prokhorenkova and Shekhovtsov(2020)]%
        {prokhorenkova2020graph}
\bibfield{author}{\bibinfo{person}{Liudmila Prokhorenkova} {and} \bibinfo{person}{Aleksandr Shekhovtsov}.} \bibinfo{year}{2020}\natexlab{}.
\newblock \showarticletitle{Graph-based nearest neighbor search: From practice to theory}. In \bibinfo{booktitle}{\emph{International Conference on Machine Learning}}. PMLR, \bibinfo{pages}{7803--7813}.
\newblock


\bibitem[Radford et~al\mbox{.}(2021)]%
        {radford2021learning}
\bibfield{author}{\bibinfo{person}{Alec Radford}, \bibinfo{person}{Jong~Wook Kim}, \bibinfo{person}{Chris Hallacy}, \bibinfo{person}{Aditya Ramesh}, \bibinfo{person}{Gabriel Goh}, \bibinfo{person}{Sandhini Agarwal}, \bibinfo{person}{Girish Sastry}, \bibinfo{person}{Amanda Askell}, \bibinfo{person}{Pamela Mishkin}, \bibinfo{person}{Jack Clark}, {et~al\mbox{.}}} \bibinfo{year}{2021}\natexlab{}.
\newblock \showarticletitle{Learning transferable visual models from natural language supervision}. In \bibinfo{booktitle}{\emph{International conference on machine learning}}. PmLR, \bibinfo{pages}{8748--8763}.
\newblock


\bibitem[Raynal(2013)]%
        {termination_detection}
\bibfield{author}{\bibinfo{person}{Michel Raynal}.} \bibinfo{year}{2013}\natexlab{}.
\newblock \bibinfo{booktitle}{\emph{Distributed Termination Detection}}.
\newblock \bibinfo{publisher}{Springer Berlin Heidelberg}, \bibinfo{address}{Berlin, Heidelberg}, \bibinfo{pages}{367--399}.
\newblock
\showISBNx{978-3-642-38123-2}
\href{https://doi.org/10.1007/978-3-642-38123-2_14}{doi:\nolinkurl{10.1007/978-3-642-38123-2_14}}


\bibitem[Ren et~al\mbox{.}(2020)]%
        {hmanns}
\bibfield{author}{\bibinfo{person}{Jie Ren}, \bibinfo{person}{Minjia Zhang}, {and} \bibinfo{person}{Dong Li}.} \bibinfo{year}{2020}\natexlab{}.
\newblock \showarticletitle{HM-ANN: efficient billion-point nearest neighbor search on heterogeneous memory}. In \bibinfo{booktitle}{\emph{Proceedings of the 34th International Conference on Neural Information Processing Systems}} (Vancouver, BC, Canada) \emph{(\bibinfo{series}{NIPS '20})}. \bibinfo{publisher}{Curran Associates Inc.}, \bibinfo{address}{Red Hook, NY, USA}, Article \bibinfo{articleno}{895}, \bibinfo{numpages}{13}~pages.
\newblock
\showISBNx{9781713829546}


\bibitem[Schuhmann et~al\mbox{.}(2021)]%
        {schuhmann2021laion}
\bibfield{author}{\bibinfo{person}{Christoph Schuhmann}, \bibinfo{person}{Richard Vencu}, \bibinfo{person}{Romain Beaumont}, \bibinfo{person}{Robert Kaczmarczyk}, \bibinfo{person}{Clayton Mullis}, \bibinfo{person}{Aarush Katta}, \bibinfo{person}{Theo Coombes}, \bibinfo{person}{Jenia Jitsev}, {and} \bibinfo{person}{Aran Komatsuzaki}.} \bibinfo{year}{2021}\natexlab{}.
\newblock \showarticletitle{Laion-400m: Open dataset of clip-filtered 400 million image-text pairs}.
\newblock \bibinfo{journal}{\emph{arXiv preprint arXiv:2111.02114}} (\bibinfo{year}{2021}).
\newblock


\bibitem[Sch{\"u}tze et~al\mbox{.}(2022)]%
        {schutze2022nearest}
\bibfield{author}{\bibinfo{person}{Konstantin Sch{\"u}tze}, \bibinfo{person}{Michael Heinzinger}, \bibinfo{person}{Martin Steinegger}, {and} \bibinfo{person}{Burkhard Rost}.} \bibinfo{year}{2022}\natexlab{}.
\newblock \showarticletitle{Nearest neighbor search on embeddings rapidly identifies distant protein relations}.
\newblock \bibinfo{journal}{\emph{Frontiers in Bioinformatics}}  \bibinfo{volume}{2} (\bibinfo{year}{2022}), \bibinfo{pages}{1033775}.
\newblock


\bibitem[Shao et~al\mbox{.}(2024)]%
        {shao2024scaling}
\bibfield{author}{\bibinfo{person}{Rulin Shao}, \bibinfo{person}{Jacqueline He}, \bibinfo{person}{Akari Asai}, \bibinfo{person}{Weijia Shi}, \bibinfo{person}{Tim Dettmers}, \bibinfo{person}{Sewon Min}, \bibinfo{person}{Luke Zettlemoyer}, {and} \bibinfo{person}{Pang Wei~W Koh}.} \bibinfo{year}{2024}\natexlab{}.
\newblock \showarticletitle{Scaling retrieval-based language models with a trillion-token datastore}.
\newblock \bibinfo{journal}{\emph{Advances in Neural Information Processing Systems}}  \bibinfo{volume}{37} (\bibinfo{year}{2024}), \bibinfo{pages}{91260--91299}.
\newblock


\bibitem[Shi et~al\mbox{.}(2016)]%
        {wukong}
\bibfield{author}{\bibinfo{person}{Jiaxin Shi}, \bibinfo{person}{Youyang Yao}, \bibinfo{person}{Rong Chen}, \bibinfo{person}{Haibo Chen}, {and} \bibinfo{person}{Feifei Li}.} \bibinfo{year}{2016}\natexlab{}.
\newblock \showarticletitle{Fast and concurrent RDF queries with RDMA-based distributed graph exploration}. In \bibinfo{booktitle}{\emph{Proceedings of the 12th USENIX Conference on Operating Systems Design and Implementation}} (Savannah, GA, USA) \emph{(\bibinfo{series}{OSDI'16})}. \bibinfo{publisher}{USENIX Association}, \bibinfo{address}{USA}, \bibinfo{pages}{317–332}.
\newblock
\showISBNx{9781931971331}


\bibitem[Simhadri et~al\mbox{.}(2022)]%
        {simhadri2022results}
\bibfield{author}{\bibinfo{person}{Harsha~Vardhan Simhadri}, \bibinfo{person}{George Williams}, \bibinfo{person}{Martin Aum{\"u}ller}, \bibinfo{person}{Matthijs Douze}, \bibinfo{person}{Artem Babenko}, \bibinfo{person}{Dmitry Baranchuk}, \bibinfo{person}{Qi Chen}, \bibinfo{person}{Lucas Hosseini}, \bibinfo{person}{Ravishankar Krishnaswamny}, \bibinfo{person}{Gopal Srinivasa}, {et~al\mbox{.}}} \bibinfo{year}{2022}\natexlab{}.
\newblock \showarticletitle{Results of the NeurIPS’21 challenge on billion-scale approximate nearest neighbor search}. In \bibinfo{booktitle}{\emph{NeurIPS 2021 Competitions and Demonstrations Track}}. PMLR, \bibinfo{pages}{177--189}.
\newblock


\bibitem[Subramanya et~al\mbox{.}(2019)]%
        {diskann}
\bibfield{author}{\bibinfo{person}{Suhas~Jayaram Subramanya}, \bibinfo{person}{Devvrit}, \bibinfo{person}{Rohan Kadekodi}, \bibinfo{person}{Ravishankar Krishaswamy}, {and} \bibinfo{person}{Harsha~Vardhan Simhadri}.} \bibinfo{year}{2019}\natexlab{}.
\newblock \bibinfo{booktitle}{\emph{DiskANN: fast accurate billion-point nearest neighbor search on a single node}}.
\newblock \bibinfo{publisher}{Curran Associates Inc.}, \bibinfo{address}{Red Hook, NY, USA}.
\newblock


\bibitem[Tellez et~al\mbox{.}(2024)]%
        {tellez2024overview}
\bibfield{author}{\bibinfo{person}{Eric~S Tellez}, \bibinfo{person}{Martin Aum{\"u}ller}, {and} \bibinfo{person}{Vladimir Mic}.} \bibinfo{year}{2024}\natexlab{}.
\newblock \showarticletitle{Overview of the SISAP 2024 Indexing Challenge}. In \bibinfo{booktitle}{\emph{International Conference on Similarity Search and Applications}}. Springer, \bibinfo{pages}{255--265}.
\newblock


\bibitem[Tian et~al\mbox{.}(2024)]%
        {smartssd}
\bibfield{author}{\bibinfo{person}{Bing Tian}, \bibinfo{person}{Haikun Liu}, \bibinfo{person}{Zhuohui Duan}, \bibinfo{person}{Xiaofei Liao}, \bibinfo{person}{Hai Jin}, {and} \bibinfo{person}{Yu Zhang}.} \bibinfo{year}{2024}\natexlab{}.
\newblock \showarticletitle{Scalable Billion-point Approximate Nearest Neighbor Search Using {SmartSSDs}}. In \bibinfo{booktitle}{\emph{2024 USENIX Annual Technical Conference (USENIX ATC 24)}}. \bibinfo{publisher}{USENIX Association}, \bibinfo{address}{Santa Clara, CA}, \bibinfo{pages}{1135--1150}.
\newblock
\showISBNx{978-1-939133-41-0}
\urldef\tempurl%
\url{https://www.usenix.org/conference/atc24/presentation/tian}
\showURL{%
\tempurl}


\bibitem[Van~Gysel et~al\mbox{.}(2016)]%
        {van2016learning}
\bibfield{author}{\bibinfo{person}{Christophe Van~Gysel}, \bibinfo{person}{Maarten de Rijke}, {and} \bibinfo{person}{Evangelos Kanoulas}.} \bibinfo{year}{2016}\natexlab{}.
\newblock \showarticletitle{Learning latent vector spaces for product search}. In \bibinfo{booktitle}{\emph{Proceedings of the 25th ACM international on conference on information and knowledge management}}. \bibinfo{pages}{165--174}.
\newblock


\bibitem[Wang et~al\mbox{.}(2021c)]%
        {milvus}
\bibfield{author}{\bibinfo{person}{Jianguo Wang}, \bibinfo{person}{Xiaomeng Yi}, \bibinfo{person}{Rentong Guo}, \bibinfo{person}{Hai Jin}, \bibinfo{person}{Peng Xu}, \bibinfo{person}{Shengjun Li}, \bibinfo{person}{Xiangyu Wang}, \bibinfo{person}{Xiangzhou Guo}, \bibinfo{person}{Chengming Li}, \bibinfo{person}{Xiaohai Xu}, \bibinfo{person}{Kun Yu}, \bibinfo{person}{Yuxing Yuan}, \bibinfo{person}{Yinghao Zou}, \bibinfo{person}{Jiquan Long}, \bibinfo{person}{Yudong Cai}, \bibinfo{person}{Zhenxiang Li}, \bibinfo{person}{Zhifeng Zhang}, \bibinfo{person}{Yihua Mo}, \bibinfo{person}{Jun Gu}, \bibinfo{person}{Ruiyi Jiang}, \bibinfo{person}{Yi Wei}, {and} \bibinfo{person}{Charles Xie}.} \bibinfo{year}{2021}\natexlab{c}.
\newblock \showarticletitle{Milvus: A Purpose-Built Vector Data Management System}. In \bibinfo{booktitle}{\emph{Proceedings of the 2021 International Conference on Management of Data}} (Virtual Event, China) \emph{(\bibinfo{series}{SIGMOD '21})}. \bibinfo{publisher}{Association for Computing Machinery}, \bibinfo{address}{New York, NY, USA}, \bibinfo{pages}{2614–2627}.
\newblock
\showISBNx{9781450383431}
\href{https://doi.org/10.1145/3448016.3457550}{doi:\nolinkurl{10.1145/3448016.3457550}}


\bibitem[Wang et~al\mbox{.}(2024)]%
        {starling}
\bibfield{author}{\bibinfo{person}{Mengzhao Wang}, \bibinfo{person}{Weizhi Xu}, \bibinfo{person}{Xiaomeng Yi}, \bibinfo{person}{Songlin Wu}, \bibinfo{person}{Zhangyang Peng}, \bibinfo{person}{Xiangyu Ke}, \bibinfo{person}{Yunjun Gao}, \bibinfo{person}{Xiaoliang Xu}, \bibinfo{person}{Rentong Guo}, {and} \bibinfo{person}{Charles Xie}.} \bibinfo{year}{2024}\natexlab{}.
\newblock \showarticletitle{Starling: An I/O-Efficient Disk-Resident Graph Index Framework for High-Dimensional Vector Similarity Search on Data Segment}.
\newblock \bibinfo{journal}{\emph{Proc. ACM Manag. Data}} \bibinfo{volume}{2}, \bibinfo{number}{1}, Article \bibinfo{articleno}{14} (\bibinfo{date}{March} \bibinfo{year}{2024}), \bibinfo{numpages}{27}~pages.
\newblock
\href{https://doi.org/10.1145/3639269}{doi:\nolinkurl{10.1145/3639269}}


\bibitem[Wang et~al\mbox{.}(2021a)]%
        {wang2021comprehensive}
\bibfield{author}{\bibinfo{person}{Mengzhao Wang}, \bibinfo{person}{Xiaoliang Xu}, \bibinfo{person}{Qiang Yue}, {and} \bibinfo{person}{Yuxiang Wang}.} \bibinfo{year}{2021}\natexlab{a}.
\newblock \showarticletitle{A comprehensive survey and experimental comparison of graph-based approximate nearest neighbor search}.
\newblock \bibinfo{journal}{\emph{arXiv preprint arXiv:2101.12631}} (\bibinfo{year}{2021}).
\newblock


\bibitem[Wang et~al\mbox{.}(2021b)]%
        {anns_survey}
\bibfield{author}{\bibinfo{person}{Mengzhao Wang}, \bibinfo{person}{Xiaoliang Xu}, \bibinfo{person}{Qiang Yue}, {and} \bibinfo{person}{Yuxiang Wang}.} \bibinfo{year}{2021}\natexlab{b}.
\newblock \showarticletitle{A comprehensive survey and experimental comparison of graph-based approximate nearest neighbor search}.
\newblock \bibinfo{journal}{\emph{Proc. VLDB Endow.}} \bibinfo{volume}{14}, \bibinfo{number}{11} (\bibinfo{date}{July} \bibinfo{year}{2021}), \bibinfo{pages}{1964–1978}.
\newblock
\showISSN{2150-8097}
\href{https://doi.org/10.14778/3476249.3476255}{doi:\nolinkurl{10.14778/3476249.3476255}}


\bibitem[Wei et~al\mbox{.}(2020)]%
        {AnalyticDB}
\bibfield{author}{\bibinfo{person}{Chuangxian Wei}, \bibinfo{person}{Bin Wu}, \bibinfo{person}{Sheng Wang}, \bibinfo{person}{Renjie Lou}, \bibinfo{person}{Chaoqun Zhan}, \bibinfo{person}{Feifei Li}, {and} \bibinfo{person}{Yuanzhe Cai}.} \bibinfo{year}{2020}\natexlab{}.
\newblock \showarticletitle{AnalyticDB-V: a hybrid analytical engine towards query fusion for structured and unstructured data}.
\newblock \bibinfo{journal}{\emph{Proc. VLDB Endow.}} \bibinfo{volume}{13}, \bibinfo{number}{12} (\bibinfo{date}{Aug.} \bibinfo{year}{2020}), \bibinfo{pages}{3152–3165}.
\newblock
\showISSN{2150-8097}
\href{https://doi.org/10.14778/3415478.3415541}{doi:\nolinkurl{10.14778/3415478.3415541}}


\bibitem[Wei et~al\mbox{.}(2018)]%
        {rdma2}
\bibfield{author}{\bibinfo{person}{Xingda Wei}, \bibinfo{person}{Zhiyuan Dong}, \bibinfo{person}{Rong Chen}, {and} \bibinfo{person}{Haibo Chen}.} \bibinfo{year}{2018}\natexlab{}.
\newblock \showarticletitle{Deconstructing {RDMA-enabled} Distributed Transactions: Hybrid is Better!}. In \bibinfo{booktitle}{\emph{13th USENIX Symposium on Operating Systems Design and Implementation (OSDI 18)}}. \bibinfo{publisher}{USENIX Association}, \bibinfo{address}{Carlsbad, CA}, \bibinfo{pages}{233--251}.
\newblock
\showISBNx{978-1-939133-08-3}
\urldef\tempurl%
\url{https://www.usenix.org/conference/osdi18/presentation/wei}
\showURL{%
\tempurl}


\bibitem[Xu et~al\mbox{.}(2023)]%
        {spfresh}
\bibfield{author}{\bibinfo{person}{Yuming Xu}, \bibinfo{person}{Hengyu Liang}, \bibinfo{person}{Jin Li}, \bibinfo{person}{Shuotao Xu}, \bibinfo{person}{Qi Chen}, \bibinfo{person}{Qianxi Zhang}, \bibinfo{person}{Cheng Li}, \bibinfo{person}{Ziyue Yang}, \bibinfo{person}{Fan Yang}, \bibinfo{person}{Yuqing Yang}, \bibinfo{person}{Peng Cheng}, {and} \bibinfo{person}{Mao Yang}.} \bibinfo{year}{2023}\natexlab{}.
\newblock \showarticletitle{SPFresh: Incremental In-Place Update for Billion-Scale Vector Search}. In \bibinfo{booktitle}{\emph{Proceedings of the 29th Symposium on Operating Systems Principles}} (Koblenz, Germany) \emph{(\bibinfo{series}{SOSP '23})}. \bibinfo{publisher}{Association for Computing Machinery}, \bibinfo{address}{New York, NY, USA}, \bibinfo{pages}{545–561}.
\newblock
\showISBNx{9798400702297}
\href{https://doi.org/10.1145/3600006.3613166}{doi:\nolinkurl{10.1145/3600006.3613166}}


\bibitem[Zeng et~al\mbox{.}(2023)]%
        {dfgasdistrifpga}
\bibfield{author}{\bibinfo{person}{Shulin Zeng}, \bibinfo{person}{Zhenhua Zhu}, \bibinfo{person}{Jun Liu}, \bibinfo{person}{Haoyu Zhang}, \bibinfo{person}{Guohao Dai}, \bibinfo{person}{Zixuan Zhou}, \bibinfo{person}{Shuangchen Li}, \bibinfo{person}{Xuefei Ning}, \bibinfo{person}{Yuan Xie}, \bibinfo{person}{Huazhong Yang}, {and} \bibinfo{person}{Yu Wang}.} \bibinfo{year}{2023}\natexlab{}.
\newblock \showarticletitle{DF-GAS: a Distributed FPGA-as-a-Service Architecture towards Billion-Scale Graph-based Approximate Nearest Neighbor Search}. In \bibinfo{booktitle}{\emph{Proceedings of the 56th Annual IEEE/ACM International Symposium on Microarchitecture}} (Toronto, ON, Canada) \emph{(\bibinfo{series}{MICRO '23})}. \bibinfo{publisher}{Association for Computing Machinery}, \bibinfo{address}{New York, NY, USA}, \bibinfo{pages}{283–296}.
\newblock
\showISBNx{9798400703294}
\href{https://doi.org/10.1145/3613424.3614292}{doi:\nolinkurl{10.1145/3613424.3614292}}


\bibitem[Zhang et~al\mbox{.}(2018)]%
        {zhang2018visual}
\bibfield{author}{\bibinfo{person}{Yanhao Zhang}, \bibinfo{person}{Pan Pan}, \bibinfo{person}{Yun Zheng}, \bibinfo{person}{Kang Zhao}, \bibinfo{person}{Yingya Zhang}, \bibinfo{person}{Xiaofeng Ren}, {and} \bibinfo{person}{Rong Jin}.} \bibinfo{year}{2018}\natexlab{}.
\newblock \showarticletitle{Visual search at alibaba}. In \bibinfo{booktitle}{\emph{Proceedings of the 24th ACM SIGKDD international conference on knowledge discovery \& data mining}}. \bibinfo{pages}{993--1001}.
\newblock


\bibitem[Zhang et~al\mbox{.}(2023)]%
        {Auncel}
\bibfield{author}{\bibinfo{person}{Zili Zhang}, \bibinfo{person}{Chao Jin}, \bibinfo{person}{Linpeng Tang}, \bibinfo{person}{Xuanzhe Liu}, {and} \bibinfo{person}{Xin Jin}.} \bibinfo{year}{2023}\natexlab{}.
\newblock \showarticletitle{Fast, Approximate Vector Queries on Very Large Unstructured Datasets}. In \bibinfo{booktitle}{\emph{20th USENIX Symposium on Networked Systems Design and Implementation (NSDI 23)}}. \bibinfo{publisher}{USENIX Association}, \bibinfo{address}{Boston, MA}, \bibinfo{pages}{995--1011}.
\newblock
\showISBNx{978-1-939133-33-5}
\urldef\tempurl%
\url{https://www.usenix.org/conference/nsdi23/presentation/zhang-zili}
\showURL{%
\tempurl}


\bibitem[Zhang et~al\mbox{.}(2024)]%
        {rummy}
\bibfield{author}{\bibinfo{person}{Zili Zhang}, \bibinfo{person}{Fangyue Liu}, \bibinfo{person}{Gang Huang}, \bibinfo{person}{Xuanzhe Liu}, {and} \bibinfo{person}{Xin Jin}.} \bibinfo{year}{2024}\natexlab{}.
\newblock \showarticletitle{Fast Vector Query Processing for Large Datasets Beyond {GPU} Memory with Reordered Pipelining}. In \bibinfo{booktitle}{\emph{21st USENIX Symposium on Networked Systems Design and Implementation (NSDI 24)}}. \bibinfo{publisher}{USENIX Association}, \bibinfo{address}{Santa Clara, CA}, \bibinfo{pages}{23--40}.
\newblock
\showISBNx{978-1-939133-39-7}
\urldef\tempurl%
\url{https://www.usenix.org/conference/nsdi24/presentation/zhang-zili-pipelining}
\showURL{%
\tempurl}


\bibitem[Zhao et~al\mbox{.}(2024)]%
        {zhao2024retrieval}
\bibfield{author}{\bibinfo{person}{Penghao Zhao}, \bibinfo{person}{Hailin Zhang}, \bibinfo{person}{Qinhan Yu}, \bibinfo{person}{Zhengren Wang}, \bibinfo{person}{Yunteng Geng}, \bibinfo{person}{Fangcheng Fu}, \bibinfo{person}{Ling Yang}, \bibinfo{person}{Wentao Zhang}, \bibinfo{person}{Jie Jiang}, {and} \bibinfo{person}{Bin Cui}.} \bibinfo{year}{2024}\natexlab{}.
\newblock \showarticletitle{Retrieval-augmented generation for ai-generated content: A survey}.
\newblock \bibinfo{journal}{\emph{arXiv preprint arXiv:2402.19473}} (\bibinfo{year}{2024}).
\newblock


\bibitem[Zhao et~al\mbox{.}(2021)]%
        {zhao2021learning}
\bibfield{author}{\bibinfo{person}{Yifan Zhao}, \bibinfo{person}{Huiyu Cai}, \bibinfo{person}{Zuobai Zhang}, \bibinfo{person}{Jian Tang}, {and} \bibinfo{person}{Yue Li}.} \bibinfo{year}{2021}\natexlab{}.
\newblock \showarticletitle{Learning interpretable cellular and gene signature embeddings from single-cell transcriptomic data}.
\newblock \bibinfo{journal}{\emph{Nature communications}} \bibinfo{volume}{12}, \bibinfo{number}{1} (\bibinfo{year}{2021}), \bibinfo{pages}{5261}.
\newblock


\bibitem[Ziegler et~al\mbox{.}(2023)]%
        {rdma1}
\bibfield{author}{\bibinfo{person}{Tobias Ziegler}, \bibinfo{person}{Jacob Nelson-Slivon}, \bibinfo{person}{Viktor Leis}, {and} \bibinfo{person}{Carsten Binnig}.} \bibinfo{year}{2023}\natexlab{}.
\newblock \showarticletitle{Design Guidelines for Correct, Efficient, and Scalable Synchronization using One-Sided RDMA}.
\newblock  \bibinfo{volume}{1}, \bibinfo{number}{2}, Article \bibinfo{articleno}{131} (\bibinfo{date}{June} \bibinfo{year}{2023}), \bibinfo{numpages}{26}~pages.
\newblock
\href{https://doi.org/10.1145/3589276}{doi:\nolinkurl{10.1145/3589276}}


\end{thebibliography}

\appendix

\end{document}